\documentclass[12pt]{jpp}
\usepackage{graphicx} 
\usepackage{amsmath}
\usepackage{amssymb}
\usepackage{url}
\usepackage{musixtex}
\usepackage[left, pagewise]{lineno}
\usepackage[dvipsnames]{xcolor}

\newcommand{\Trr}{\mathrm{Tr}} 
 
\newcommand{\MM}{\mathsf{M}}
\newcommand{\trm}{\Trr(\MM)}

\newcommand{\BBB}{\mathbf{B}}

\newcommand{\ipca}{I_\text{PC A}}
\newcommand{\ipcb}{I_\text{PC B}}
\newcommand{\inpc}[1]{I_\text{NPC #1}}
\newcommand{\icc}{I_\text{CC}}
\newcommand{\imain}[1]{I_\text{main #1}}
\newcommand{\iaux}[1]{I_\text{aux. #1}}

\newcommand{\dmin}{d_\text{min}}
\newcommand{\dmax}{d_\text{max}}
\newcommand{\dqfm}{d_\text{QFM}}
\newcommand{\psol}{P_\text{SOL}}

\begin{document}
\title[Giant islands in HSX and W7-X]{Computational studies of giant edge islands and unpaired X-points in HSX and W7-X by manipulating coil currents}
\author{R. Davies\aff{1}\corresp{\email{robert.davies@ipp.mpg.de}}, L. Day\aff{2}, L. Sanchez-Garay\aff{2}, D. Boeyaert\aff{3}, A. Wolfmeister\aff{3}, B. Geiger\aff{3}, G. F. Harrer\aff{4}, J. Geiger\aff{1}, the W7-X team}
\affiliation{${}^1$Max Planck Institute for Plasma Physics, Wendelsteinstraße 1, 17491 Greifswald, Germany
${}^2$Imperial College London, London, UK
${}^3$University of Wisconsin-Madison, Madison, USA
${}^4$Hampton University, Virginia, USA}
\date{\today}
\maketitle

\begin{abstract}
We present magnetic configurations in the Helically Symmetric eXperiment (HSX)
and Wendelstein 7-X (W7-X), in which the edge magnetic structure exhibits
island chains which are spatially larger than the previously reported configurations.
These ``giant" island chains (with rotational transform $\iota=4/3$ or $4/4$ for HSX and $\iota=5/6$, $5/5$ or $5/4$ for W7-X) are obtained by reducing the coil current in main coil 6 for HSX and non-planar coil 5 for W7-X (i.e. the coil nearest the up-down symmetric cross-section $\phi=36^\circ$ for W7-X and $\phi=45^\circ$ for HSX). This appears a sufficient (but not necessary) condition for giant islands. The giant islands create X-point legs which transport plasma to the plasma-facing components (PFCs). In the most extreme cases, the island O-points leave the domain of the field line map and the divertor legs of the remaining ``unpaired" X-points do not close around the island. We show for W7-X, via the construction of quadratic flux-minimising surfaces, that these island size trends can mostly be interpreted as changes in magnitude of a resonant perturbation on a changing resonant ``unperturbed" magnetic surface. We use the anisotropic heat diffusion code EMC3-Lite to find ``giant island" W7-X configurations which are promising for PFC heat loads. These configurations significantly degrade the neoclassical transport performance but allow exploration of novel divertor regimes; for example, we show that the internal island pitch can be increased. This leads to favourable downstream/upstream plasma density ratios according to simple models; an attractive finding which motivates further investigation. 
\end{abstract}

\section{Introduction}\label{sec:intro}
\begin{figure}
    \centering
    \includegraphics[width=0.8\linewidth]{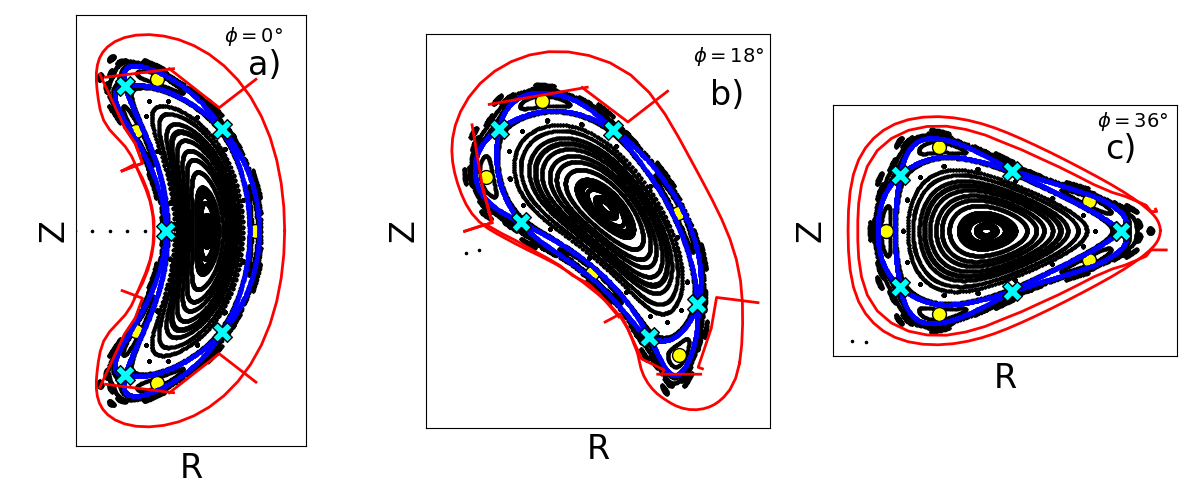}        
    \includegraphics[width=0.8\linewidth]{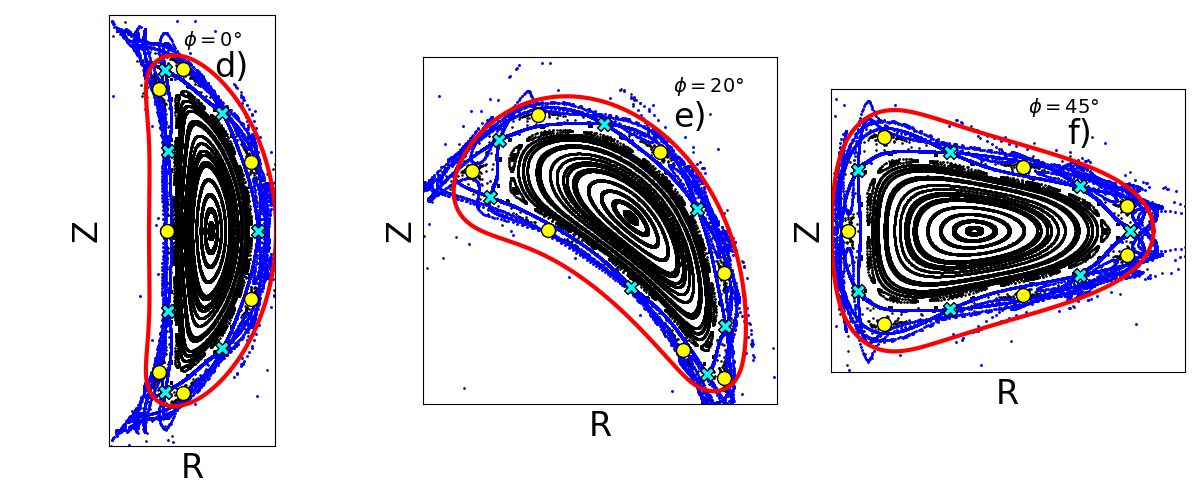}
    \caption{
     a-c) Poincaré plots and $\iota=5/5$ X/O-points at different toroidal locations for W7-X standard (``EIM") configuration. d-f) Poincaré plots and $\iota=8/7$ X/O-points at different toroidal locations for HSX standard (``QHS") configuration. Plasma-facing components are shown in red (the vessel wall for HSX, and the wall, divertor plates, baffles and heat shields for W7-X).
 }
    \label{fig:hsx_w7x_poincare}
\end{figure}
\begin{figure}
    \centering
    \includegraphics[width=0.4\linewidth]{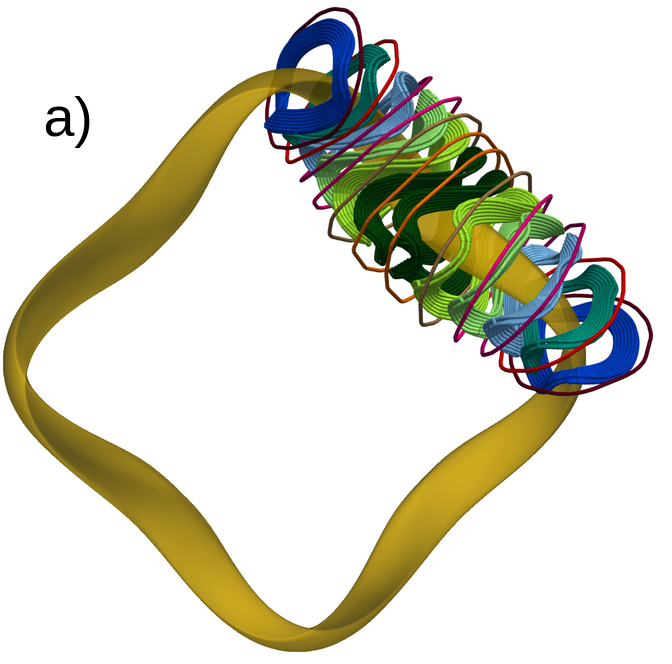}
    \includegraphics[width=0.4\linewidth]{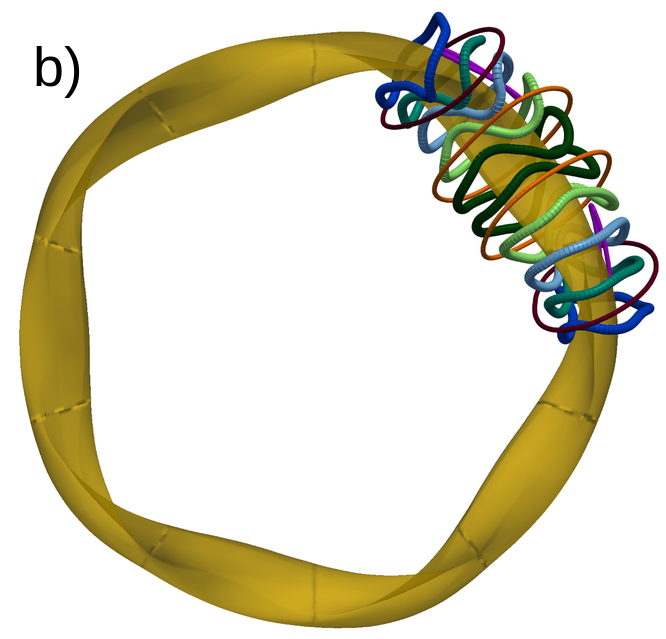} 
    \includegraphics[width=0.48\linewidth]{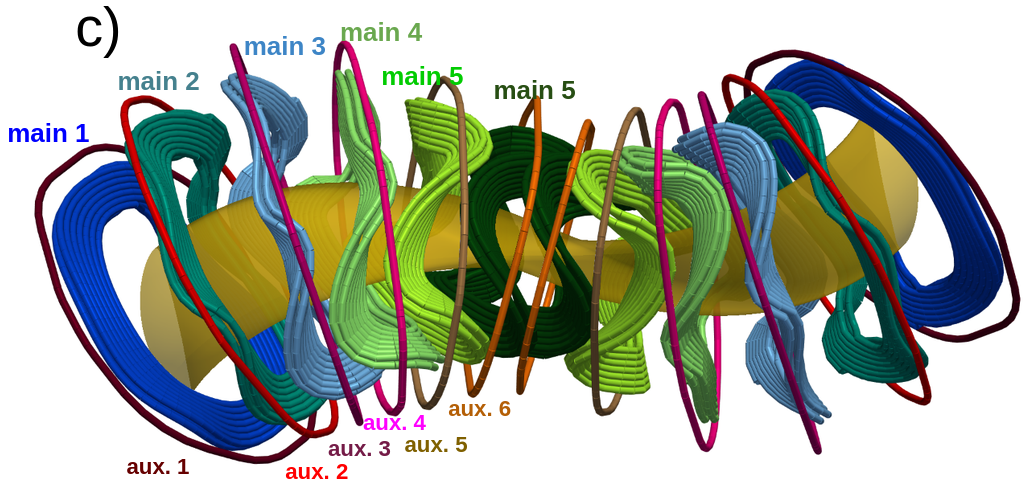}
    \includegraphics[width=0.48\linewidth]{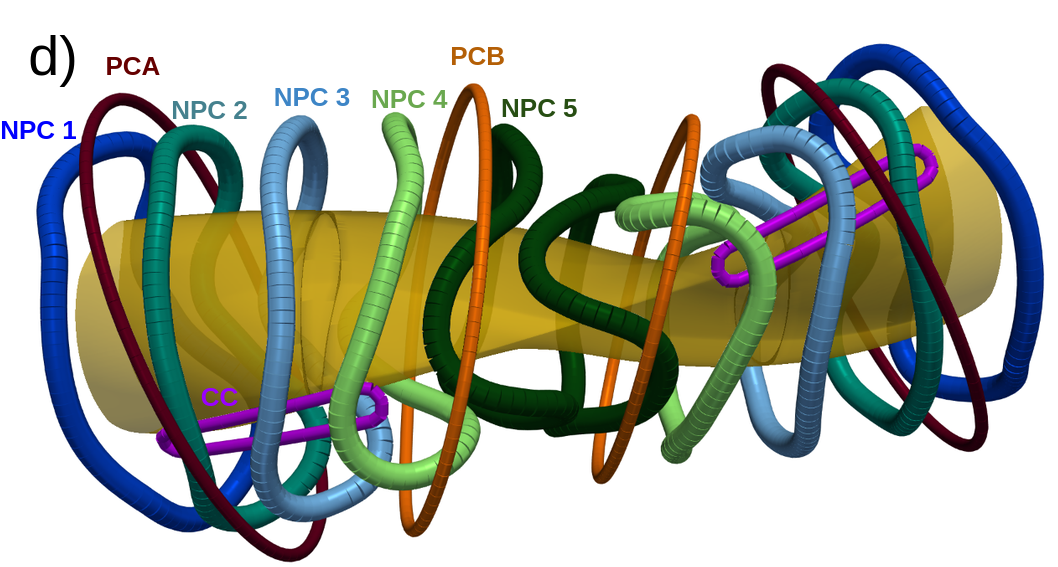}
    \caption{
    a) Top-down view of HSX plasma and coil system b) Top-down W7-X view. Coils shown for one field period. c) and d) Side-on views of a field period for HSX and W7-X. Unique coil geometries for each are colour-coded and labelled ``main 1-6" and ``aux. 1-6" for HSX and ``NPC 1-5", "PCA,B" and "CC" for W7-X (non-planar coils, planar coils and control coils respectively).
 }
    \label{fig:hsx_w7x_coils_3d}
\end{figure}
Stellarator divertor design and optimisation is of considerable importance for stellarator reactors, since reactors must suitably manage heat exhaust and the removal of helium ash and other impurities \citep{burnett1958divertor, konig2002divertor, gates2018stellarator}. A popular divertor candidate is the island divertor, used in Wendelstein 7-X (W7-X) \citep{renner2004} and incorporated into the design of several recent reactor concepts \citep{Warmer_2022, lion2025stellaris, bader2025power, goodman_squid_2025, sanchez2026ciemat}. The island divertor uses a low-order magnetic island chain to spatially separate the confined plasma from the plasma-facing components (PFCs). An island separatrix (consisting of field lines which travel from island X-point to island X-point, encircling an O-point) diverts the plasma onto PFCs, as illustrated in figure \ref{fig:hsx_w7x_poincare} a-c) for HSX and W7-X (island separatrix shown in blue). 

Two alternatives to the island divertor (selected here for their relevance to this work) are: (1) X-points which divert plasma onto PFCs, without encircling the O-point of a magnetic island chain. This is the case in many tokamaks \citep{stangeby2000tutorial}, the Large Helical Device \citep{ohyabu1992helical}, several future stellarator designs \citep{swanson2025overview, harrer2026star_lite} and several analytic stellarator equilibria/Hamiltonian systems \citep{freidberg2014ideal, solov1970plasma, davies_topology_2025}; (2) to deliberately introduce chaos into the edge magnetic field such that X- and O-points, whilst still present, do not divert the plasma onto PFCs in a geometrically simple way but rather via complex overlapping lobe structures characteristic of Hamiltonian chaos \citep{mackay1984transport, bensimon1984extended} (an example is shown in figure \ref{fig:hsx_w7x_poincare} d-f, with chaotic lobes in blue, and the lobes are also shown in figure \ref{fig:hsx_3d_illustrations} a-b). This approach has been investigated via the Helically Symmetric eXperiment (HSX) \citep{bader2013simulations, garcia2025_hsx}, the Compact Toroidal Hybrid \citep{garcia2023exploration}, several symmetry-breaking tokamak experiments \citep{evans2005experimental, jakubowski2007observation} and some W7-X configurations \citep{smiet2025turnstile, boeyaert2025analysis}. It is worth emphasising that magnetic chaos is not mutually exclusive with either island chains or non-island diverting X-points, but rather a physical effect which can play in concert with such structures. Both of these alternatives arise within the literature on ``non-resonant divertors" (NRDs) \citep{boozer2015stellarator, bader2017hsx, punjabi2020simulation, garcia2023exploration, boeyaert2025analysis, boeyaert2025towards, davies_topology_2025, harrer2026star_lite}, although the defining features and characteristics of NRDs vary across literature.

This work explores how changes to coil currents can manipulate island divertors in two neoclassically optimised modular coil stellarators, HSX and W7-X. The remainder of this introduction describes the motivation for island size studies and then introduces HSX and W7-X. 

Island size variation has been studied in W7-X \citep{barbui2020measurements, winters2024first, feng2024conditions} as well as in simple geometries of nested circular flux surfaces \citep{feng2022review, makarov2025modeling, onorati2026investigation}. The latter shows the dependencies of island size $r_i$ and the internal field line pitch $\Theta_i$ (i.e. the rate at which field lines inside the island circle around the O-point) \citep{feng2022review}:
\begin{align}
    r_i &= 2 \sqrt{\frac{R b_{rm}}{\iota'm}}, \label{eq:island_width} \\
    \Theta_i (\Delta r, \Delta\theta) &= \frac{4}{m} \frac{r_a}{r_i}b_{rm}\sqrt{\left(\frac{m}{4} \frac{r_i}{r_a}\right)^2\sin^2(\Delta\theta)+ \left(\frac{\Delta r}{r_i} \right)^2},    
\end{align}
where $R$ is the major radius, $m$ is the poloidal mode number of the island chain, $\iota'=\text{d}\iota/\text{d}r$ is the magnetic shear at the island location (where $r$ is the minor radius), $b_{rm}$ represents the magnitude of resonant radial magnetic perturbation, $r_a$ is the minor radius of the resonant surface and $\Delta r$, $\Delta \theta$ are the radial and poloidal coordinates with respect to the O-point. Island size can thus be controlled by $b_{rm}$ and $\iota'$, with different implications for $\Theta_i$ and therefore the parallel distance to the PFCs (here we parametrise the parallel distance to PFCs by the PFC-to-PFC connection length $L_C$). $\Theta_i$ and $L_C$ then affect power spreading on PFCs and the ratio of downstream to upstream plasma density $n_d/n_u$ and temperature $T_d/T_u$ according to simple models \citep{feng2011comparison}. 

The effect of varying island size in simple geometries has also been examined computationally \citep{onorati2026investigation} using the EMC3-EIRENE code \citep{feng2014recent}. This study keeps the simple circular geometry and increases the island size either by increasing $b_{rm}$, reducing $\iota'$, or varying both to increase $r_i$ at constant $\Theta_i$. These show that $n_d/n_u$ increases with $r_i$ in all cases, provided that $n_u$ is sufficiently large. There is no single reason for this trend; the change in $L_C$, transparency of the island O-point to neutral particles and momentum loss along the power-carrying field lines are reported as significant factors. Beyond simple geometries, there is also simulation and experimental data on the effect of varying $r_i$, $\Theta_i$ and $L_C$ in W7-X \citep{barbui2020measurements, winters2024first, feng2024conditions}. It is worth noting that in W7-X neither the magnetic geometry nor the PFC geometry is simple (for example, the flux surfaces are highly non-circular, perturbations are not spectrally pure and the islands have nonuniform pitch structure). 

Another motivation for larger islands is that if divertor legs can be made long and straight, the construction of divertor and baffles might be easier for geometric reasons. The physical effect of tight baffling is not included in simple models but has been explored computationally for island divertors \citep{maaziz2025investigating, bader2025power} and has been experimentally demonstrated in LHD to improve neutral pressure in the divertor and the neutral pumping efficiency \citep{motojima2018establishment, wenzel2026ultra}. Establishing examples of ``giant islands" in low-shear stellarators such as HSX and W7-X are therefore motivated both from the simple parametric dependencies in island divertor models (for example, obtaining a better understanding of the relation between $n_d/n_u$ and $r_i$) and for their possible contribution towards ``advanced" stellarator divertor concepts. Long-legged divertors have been employed in tokamaks \citep{fishpool2013mast, gallo2018impact, sun2023performance}, often also combined with tight baffling, and have reported favourable results; increased neutral pressures in the divertor region, increased radiative cooling and volumetric recombination of the plasma and greater power spreading.

We now describe two stellarators central to this work. HSX \citep{Geiger_HSX_2026} is a four field period stellarator, with 6 main coils per half field period ($\phi=0^\circ$ to $\phi=45^\circ$) (numbered 1-6, with 1 at the bean cross section and 6 at the straight section) and 6 auxiliary coils per half field period. In each half period these coils are copied and rotated to create a periodic and stellarator-symmetric magnetic field \citep{dewar1998stellarator}. 
The coils are shown in figure \ref{fig:hsx_w7x_coils_3d} a) and c), together with a magnetic flux surface in the confined region. We label the coil currents as $\imain{1-6}$ for the main coils and $\iaux{1-6}$ for the auxiliary coils. When only the main coils are powered and carry identical current (i.e. $\imain{1}=\imain{2}=\imain{3}=\imain{4}=\imain{5}=\imain{6}$, $\iaux{1-6}=0$), they produce the ``standard" highly quasi-helically symmetric magnetic field \citep{anderson2006overview, Geiger_HSX_2026}. The on-axis rotational transform slightly exceeds $1$ and there is a (chaotic) $\iota=8/7$ island chain around the location of the vessel wall. Poincaré plots and the periodic $\iota=8/7$ field lines (X- and O-points) are shown in figure \ref{fig:hsx_w7x_poincare} d-f). 

W7-X is a stellarator optimised for neoclassical transport via (approximate) quasi-isodynamicity \citep{wolf2016wendelstein, dinklage2018magnetic, grulke2026overview}.  W7-X has five field periods, each containing five non-planar coils (labelled 1-5), two planar coils (labelled A and B) and one ``control" (or ``sweep") coil per half field period ($\phi=0^\circ$ to $\phi=36^\circ$).  These coils are shown in figure \ref{fig:hsx_w7x_coils_3d} b), d) and are stellarator-symmetric and periodic. In practice, small error fields in W7-X break the periodicity/symmetry \citep{lazerson2017error, bozhenkov2019measurements}, but in this modelling work this effect is ignored. The five trim coils used to correct these error fields \citep{rummel2011trim, rummel2013wendelstein} are not shown in figure \ref{fig:hsx_w7x_coils_3d} and are not discussed in this work. Poincaré sections for the standard configuration (for which the planar and control coils carry zero current and the non-planar coils carry identical current, i.e. $\inpc{1}=\inpc{2}=\inpc{3}=\inpc{4}=\inpc{5}$, $\ipca=\ipcb=\icc=0$) is shown in figure \ref{fig:hsx_w7x_poincare} a-c).

Our focus is the lowest order island chains which are experimentally realisable in these two devices, namely: the $\iota=4/3$ and $4/4$ chains in HSX (which can be selected by changing the auxiliary coil currents \citep{gerhardt2004measurements}) and the $\iota=5/4, 5/5, 5/6$ island chains in W7-X (which can be selected by changing the planar coil currents). These are ``natural resonances", i.e. the toroidal periodicity of the islands match the number of field periods (and thus the resonances are compatible with stellarator periodicity and symmetry). In both devices, we find empirically (via simulations) that the island size can be ``tuned" by a single parameter; the current in the coils nearest the symmetry plane in the straight section of the device i.e. $\inpc{5}$ in W7-X (which is nearest to the $\phi=36^\circ$ plane) and $\imain{6}$ in HSX, which is nearest the $\phi=45^\circ$ cross-section. The main goal of this paper is to announce the observation of giant edge structures under these conditions (an observation which has not hitherto been reported) and to examine such configurations. We stress that the core plasma in these examples is not optimised i.e. in creating these novel edge structures, we degrade the core performance. Nevertheless, such configurations might be suitable for divertor experiments, thus informing what future combined core/edge-optimised experiments should target.

This work is organised as follows. Section \ref{sec:1d_scans} describes scans of $\imain{6}$ for HSX and $\inpc{5}$ for W7-X for the different edge $\iota$ values, including two approaches to quantify the island size in sections \ref{sec:island_size_quantification_d} and \ref{sec:island_size_quantification_qfm}. This shows quantitatively how the island size grows as $\imain{6}$ and $\inpc{5}$ decreases. In sections \ref{sec:hsx_selected_configs} and \ref{sec:w7x_selected_configs} we perform more detailed analysis on several selected HSX and W7-X configurations. We also include some discussion on the experimental viability of these examples, and provide an outlook in section \ref{sec:conclusions}. 

\section{Multi-machine island size scans}\label{sec:1d_scans}
\begin{figure}
    \centering
    \includegraphics[width=0.9\linewidth]{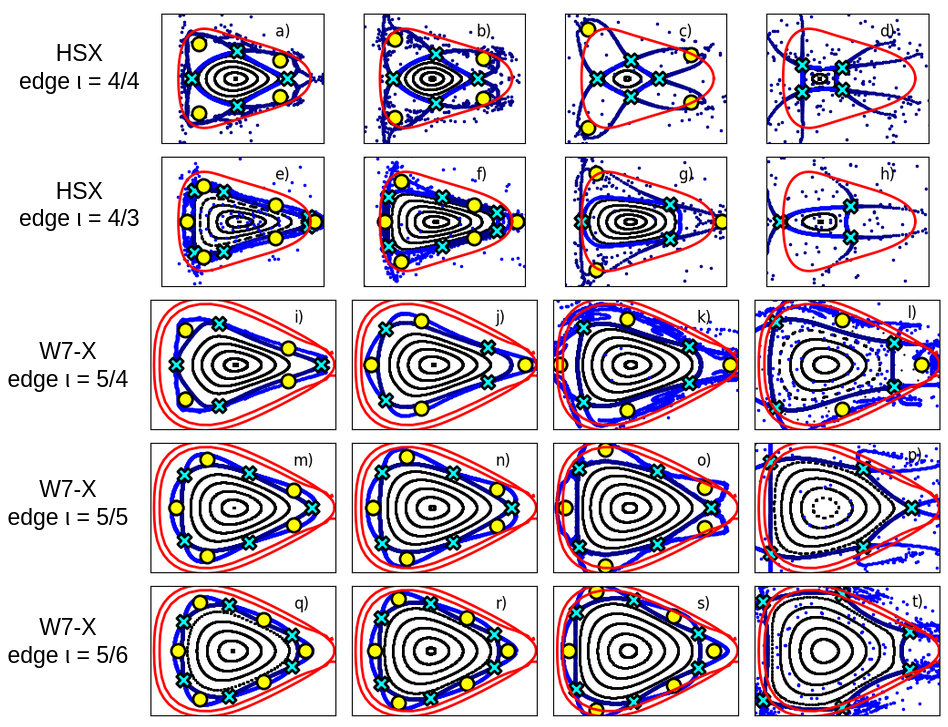}
    \caption{
    Island size variation for HSX and W7-X across five families of magnetic configurations. Rows 1: HSX, with $\imain{1-5}=1$ and $\imain{6}=(1,0.7,0.4,0.1)$ (a) - d)) (normalised currents), with edge $\iota=4/4$ selected by setting $\iaux{1-6}=0.1$. Row 2: the same but with edge $\iota=4/3$ via $\iaux{1-6}=-0.1$. Rows 3-5: W7-X, with $\inpc{1-4}=1$ and  $\inpc{5}=(1,0.7,0.4,0.1)$, for edge $\iota=5/4$ (row 3, $\ipca=\ipcb=-0.23$), $5/5$ (row 4, $\ipca=\ipcb=0$), $5/6$ (row 5, $\ipca=\ipcb=0.25$).
    }
    \label{fig:hsx_w7x_1d_scans_poincare}
\end{figure}
\begin{figure}
    \centering
    \includegraphics[width=0.9\linewidth]{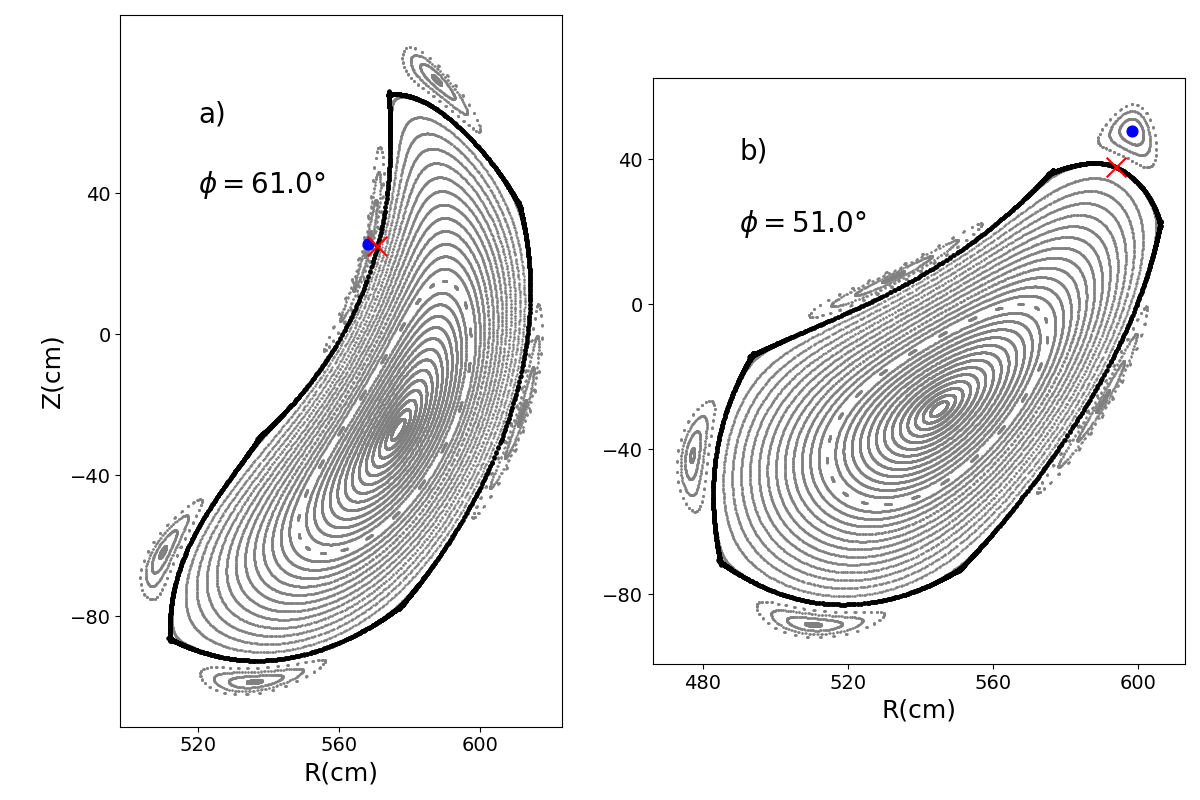}
    \caption{
    Illustration of minimum and maximum distance ($\dmin$, $\dmax$) calculation for the W7-X standard configuration. The island O-point (blue) is shown at the locations of its closest and furthest approach from the inner island separatrix (black). Poincaré data is shown in grey.
    }
    \label{fig:w7x_sepo_illustration}
\end{figure}
\begin{figure}
    \centering
    \includegraphics[width=0.9\linewidth]{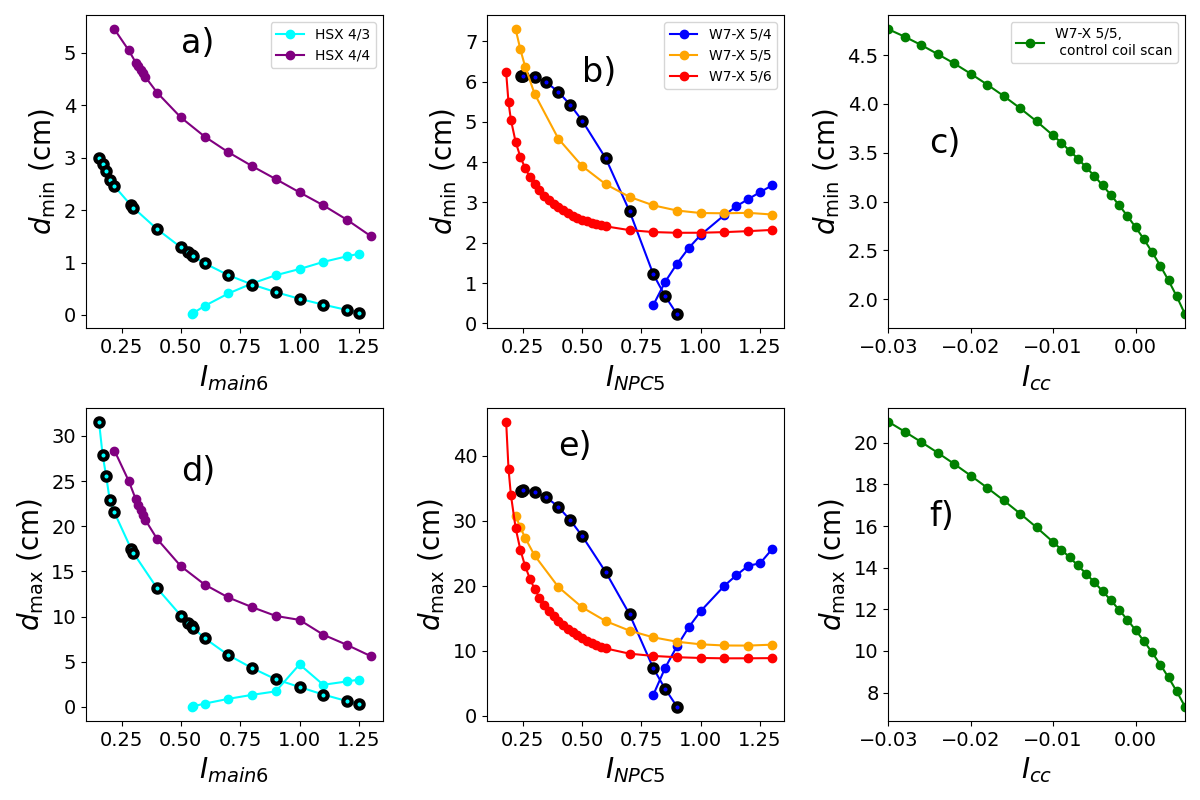}
    \caption{
    Island size as coil current scanned in HSX and W7-X. a) $\dmin$ and d) $\dmax$ for HSX for $\iota=4/4$ and $4/3$ as $\imain{6}$ is scanned. b) $\dmin$ and e) $\dmax$ for W7-X ($\iota=5/4,5/5,5/6$) as $\inpc{5}$ is scanned. c) $\dmin$ and f) $\dmax$ for W7-X ($\iota=5/5$) as $\icc$ is scanned. For HSX $\iota=4/3$ and W7-X $\iota=5/4$, two ``branches" are present (i.e. two sets of X/O points) over the scan, shown as black-outlined circles and circles without outlines.
    }
    \label{fig:w7x_hsx_sepo_calculation}
\end{figure}
For HSX, we study the island size variation as we scan $\imain{6}$. In each scan we keep $\imain{1-5}=1$ and $\iaux{1-6}$ equal and constant, with $\iaux{1-6} = \pm 0.1$ for $\iota=4/4$ or $\iota=4/3$. The currents given here are the fraction of the main coil amp-turn current. This is the same as the (normalised) current passing through the filaments shown in figure \ref{fig:hsx_w7x_coils_3d} a) and c), but with $\iaux{1-6}$ divided by 14. This is because the main coils have 14 windings and the auxiliary coils have 10 windings, but the auxiliary coils are only represented by a single winding in this work. Using the main coil amp-turn current is a convention consistent with other HSX literature \citep{bader2013simulations, bader2017hsx, garcia2025_hsx, boeyaert2025towards}. The choice to scan $\imain{6}$ rather than the other coils is driven by the observation (via simulation) of giant edge structures when $\imain{6}=0$. The effect of setting $\imain{1-5}=0$ is described in appendix \ref{app:hsx_coil_scans}. For W7-X, we scan $\inpc{5}$ while keeping $\inpc{1-4}=1$ and $\icc=0$, with the $\iota=(5/6$, $5/5$, $5/4$) resonances selected by setting $\ipca=\ipcb= (0.25 $, $0$, $-0.23$) respectively (for W7-X, these currents correspond to the normalised currents through the filaments in figure \ref{fig:hsx_w7x_coils_3d}). The filaments in figure \ref{fig:hsx_w7x_coils_3d} b) and d) represent 108 windings for the non-planar coils, 36 windings for the planar coils and 8 windings for the control coils. We describe the scans first qualitatively and then quantitatively. 

The qualitative results are shown in figure \ref{fig:hsx_w7x_1d_scans_poincare}, with each row representing a 1D scan. Rows 1 and 2 show HSX island size variation; for both $\iota$ values, the edge islands increase as $\imain{6}$ decreases.  In both cases, there comes a point where the O-point becomes sufficiently close to the coils that it no longer completes a periodic orbit, and thus the configuration ceases to resemble a typical island divertor. One difference between the HSX scans is the transition from an $n/m =8/6 \rightarrow 4/3$ island chain as $\imain{6}$ decreases, i.e. there is a transition from 6 islands to 3 (compare figure \ref{fig:hsx_w7x_1d_scans_poincare} f) and g) ). It is worth emphasising that at no point during this scan does the island chain become small, consistent with other observations of island phase transitions in W7-X \citep{davies2025characterisation} and SQuIDs \citep{davies_squid_2025}.

The results for the three W7-X scans are shown in figure \ref{fig:hsx_w7x_1d_scans_poincare} rows 3-5. Again, a common feature is island size increasing as $\inpc{5}$ decreases, with the eventual disappearance of O-points. Again, this choice is driven by empirical observation, with $\inpc{1-4}=0$ presented and discussed in  appendix \ref{app:w7x_coil_scans}. The $\iota=5/4$ cases also exhibits an island phase transition, creating two ``branches" of island behaviour; one set of islands becomes larger as $\inpc{5}$ increases, another set becoming larger as $\inpc{5}$ decreases. 

It is worth noting that these configurations are not necessarily experimentally achievable, and are not necessarily island divertors; the changing shape of the confined region can cause the nested surfaces to limit on the PFCs, with the edge structures described here becoming irrelevant for divertor performance. In other words, the divertor properties are a function of both the island size (discussed in this section) \textit{and} the location of the PFCs with respect to these islands and nested flux surfaces (which is considered in sections \ref{sec:hsx_selected_configs} and \ref{sec:w7x_selected_configs}). 

We now present two methods to quantify the island size variation across these configuration families. The first (section \ref{sec:island_size_quantification_d}) characterises the island by a minimum/maximum distance of the island O-point from the inner separatrix. The second (section \ref{sec:island_size_quantification_qfm}) calculates $r_i$ in equation \eqref{eq:island_width}, by deconstructing the magnetic field into an ``unperturbed" set of nested flux surfaces and a resonant perturbation. 

\subsection{Island size calculation: $\dmin$ and $\dmax$ }\label{sec:island_size_quantification_d} 
Our first island size metric is the minimum distance from the inner separatrix of the island to the island O-point, $\dmin$ (which is very similar to the $\text{min}(\delta_N)$ metric used by \cite{garcia2025_hsx}). With the inner separatrix serving as a proxy for the last closed magnetic surface, this provides a minimum radial extent of the divertor legs; the divertor legs cannot close without encircling the O-point, so this sets the minimum island size. The second metric is the maximum distance from the inner separatrix to the O-point, $\dmax$; the divertor legs must extend at least this far at one location. An example of the $\dmin$ and $\dmax$ calculation is shown in figure \ref{fig:w7x_sepo_illustration}. $\dmin$ and $\dmax$ are PFC-agnostic descriptions of magnetic structure, and are only divertor-relevant when the island is intersected by the PFCs; if (for example) the inner separatrix of the island chain intersects the PFCs then $\dmin$ and $\dmax$ are still well-defined but the stellarator acts as a limiter rather than an island divertor.

This approach is implemented, locating X/O points with the same root-finding scheme as presented in \cite{davies2025characterisation}. We also use the finite-differencing scheme described here to calculate the Jacobian of the field line map $\textbf{M}$, which reveals the ``flavour" of the periodic orbit ($\trm>2$ for X-points and $-2 < \trm<2$ for O-points). The separatrix is found by sampling points on the eigenvectors of $\textbf{M}$ of the X-points and tracing these field lines for a finite number of field periods, selected so that the inner separatrix is well-sampled but before the chaotic lobes become warped and stretched around the islands.

$\dmin$ and $\dmax$ are shown for HSX in figure \ref{fig:w7x_hsx_sepo_calculation} a) and d). This shows the island size, by either characterisation,  increasing as $\imain{6}$ decreases. It also reveals that $\dmin$ and $\dmax$ are larger for the $\iota=4/4$ chain than $\iota=4/3$ chain for constant $\imain{6}$. For $\iota=4/3$ there are two ``branches" of island size, corresponding to the two island phases (at a given $\dmin$, when both phases exist then there is an $8/6$ island chain i.e. six O-points and six X-points). $\dmin(\imain{6})$ and $\dmax(\imain{6})$ show one branch (black-outlined circles) growing as $\imain{6}$ decreases and the other branch (non-outlined circles) shrinking as $\imain{6}$ decreases. Finally, we note that $\dmin$ is smaller than $\dmax$ by a factor of 3-10; the physical island width varies strongly along the trajectory of the O-point.

Similar trends for W7-X are shown in figure \ref{fig:w7x_hsx_sepo_calculation} b) and e). For $\iota=5/5$ and $5/6$, a single island chain is present and grows sharply as $\inpc{5}$ is reduced, with $\dmin$ and $\dmax$ behaving similarly. Both increase by a factor of approximately $3$ over the scanned range. The edge $\iota=5/4$ chain displays two sets of O-points relating to the two island phases, with one (the ``normal" $5/4$ phase) sharply decreasing to zero as $\inpc{5}$ decreases and the other increasing. A curious difference is that $\dmin$ and $\dmax$ increases more slowly as $\inpc{5}$ decreases at low $\inpc{5}$. 

Finally, it is also instructive to compare scanning $\inpc{5}$ in W7-X with another methodology to adjust island size, $\icc$. The result of such a scan is shown in figure \ref{fig:w7x_hsx_sepo_calculation} c) and f). Comparing with the $\inpc{5}$ scan with edge $\iota=5/5$, we find that $\icc$ has a far greater effect on $\trm$, $\dmin$ and $\dmax$ for a given current. However, because $\inpc{5}$ has a much greater range, $\dmin$ and $\dmax$ can be made larger.

\subsection{Island size calculation via QFM surface calculation}\label{sec:island_size_quantification_qfm} 
\begin{figure}
    \centering
    \includegraphics[width=0.8\linewidth]{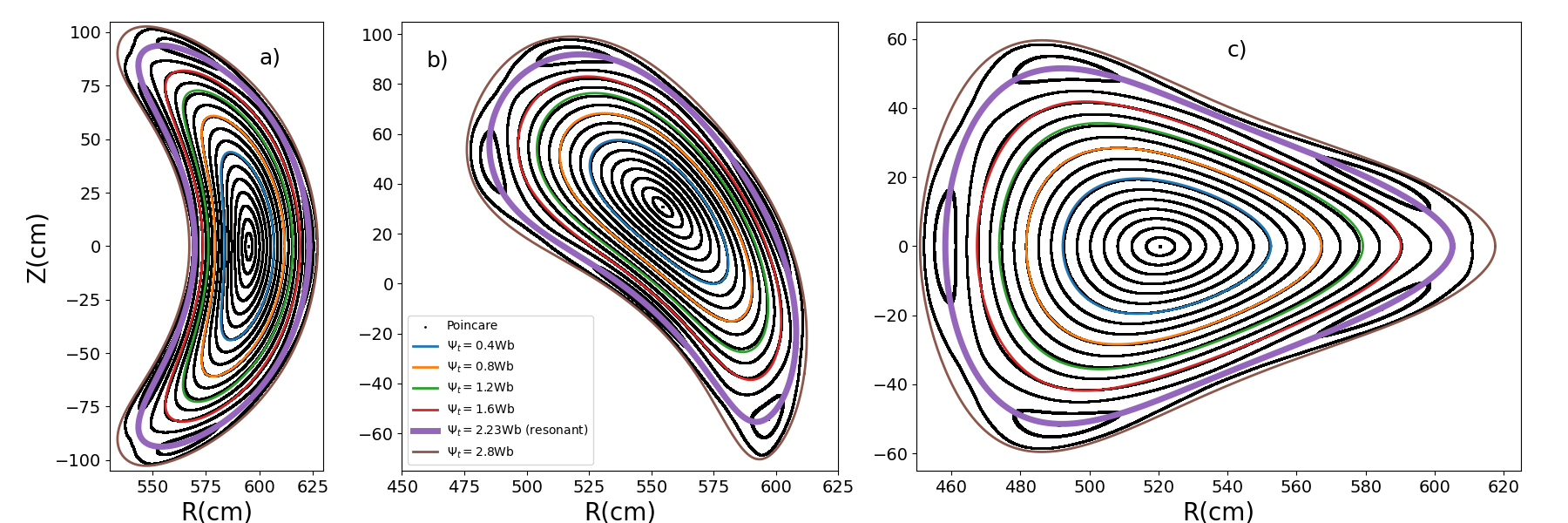}
    \includegraphics[width=0.8\linewidth]{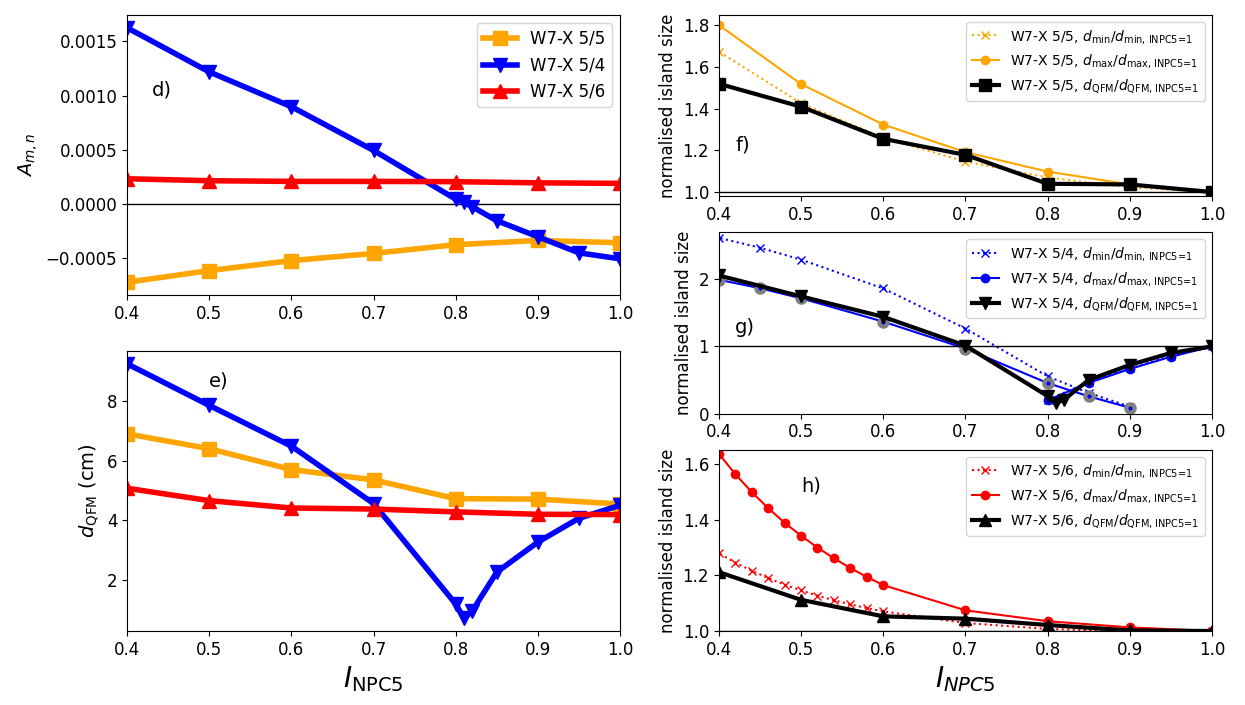}
    \caption{
    QFM calculations for W7-X. a-c) Poincaré surfaces calculated using \texttt{simsopt}'s field line tracer for the W7-X standard configuration, and the QFM surfaces shown as coloured surfaces. The QFM surface passing through the island X- and O-points (thicker lilac line) is used to estimate the island width $\dqfm$. 
    d) Resonant perturbation $A_{m,n}$ for the $\inpc{5}$ scans for $\iota=5/5, 5/4$ and $5/6$. 
    e) Island radius $\dqfm$ for $\inpc{5}$ scans calculated using equation \eqref{eq:island_width}. 
    f-h) $\dmin, \dmax$ and $\dqfm$ for the $\inpc{5}$ scans, normalised to their values at $\inpc{5}=1$. 
    }
    \label{fig:qfm_calculations}
\end{figure}
An alternative method to calculate the island size is to use the simple island model presented in section \ref{sec:intro}. As discussed, this applies several simplifying assumptions, but provides physics insight by showing how changes to the radial resonant perturbation $b_{rm}$ and magnetic shear $\iota'$ contribute to the changing island size. We apply this methodology to the W7-X scans shown in section \ref{sec:1d_scans}.

Applying equation \eqref{eq:island_width} to calculate the island radius $r_i$ requires the major radius of the resonant surface $R$, the poloidal mode number $m$, $b_{rm}$ and $\iota'$. We here take $R=5.5$m as constant and set $m=4,5,6$ for the three scans. It then simply remains to calculate $\iota'$ and $b_{rm}$ as the magnetic configuration changes.

$\iota'$ and $b_{rm}$ are calculated by computing Quadratic Flux Minimising (QFM) surfaces \citep{dewar1994almost, hudson2009ghost} for the candidate magnetic configurations. QFM surfaces are toroidal surfaces which, for a targeted enclosed toroidal magnetic flux, minimise the quantity:
\begin{equation}
    f(S) = \frac{\oint_S \text{d}A\ (\mathbf{B \cdot \hat{n})^2}}{\oint_S \text{d}A\ |\mathbf{B}|^2},
\end{equation}
where $\hat{n}$ is the unit vector normal to the surface, $\text{d}A$ is a surface element and the integral is taken over the surface. We compute these surfaces using the \texttt{simsopt} package \citep{landreman2021simsopt} and the code to produce these is publicly available (see section \ref{sec:data_availability}). An example of several QFM surfaces for the W7-X standard configuration is shown in figure \ref{fig:qfm_calculations} a-c). It can be seen that these coincide with the Poincaré surfaces, although far from the axis we must increase the spectral resolution defining the surface to achieve this agreement.

The shear on the resonant-$\iota$ QFM surface is estimated by either interpolating or extrapolating the $\iota$ profile from good magnetic surfaces. Similarly, we define a poloidal and toroidal coordinate system on the resonant-$\iota$ surface by constructing Boozer coordinates for magnetic flux surfaces near to the resonance (via \texttt{simsopt}) and extrapolating these onto the resonant-$\iota$ QFM surface. The radial perturbation $b_r$ is calculated as the magnetic flux through the surface normalised to the toroidal magnetic field, $b_r = \mathbf{B \cdot \hat{n}}/B_\phi$. We then Fourier decompose $b_r$ as
 $b_r =\mathbf{B\cdot \hat{n}}/B_t$
\begin{equation}
    b_r = \sum_{m,n} \left[ A_{m,n} \sin(m\theta - n\phi) + B_{m,n} \cos(m\theta - n\phi)\right].
\end{equation}
The resonant $A_{m,n}$ corresponds to $b_{rm}$ in equation \eqref{eq:island_width} (and, as a consequence of stellarator symmetry, $B_{m,n}=0$). Here we only consider the lowest-order resonant contributions ($(m,n) = (5,5), (4,5), (6,5)$). $A_{m,n}$ is shown in figure \ref{fig:qfm_calculations} d). This shows the weakest variation in $A_{m,n}$ for $\iota=5/6$ and strongest variation in $\iota=5/4$, with $A_{m,n}$ crossing zero to become large and positive as $\inpc5$ is reduced. Using this information, we calculate the island width, which we label $d_\text{QFM}$ (which is identical to $r_i$ in \eqref{eq:island_width}). The values of $\dqfm$ fall between the $\dmin$ and $\dmax$ values.

To compare this with the $\dmin$ and $\dmax$ metrics we plot the island widths for all three, normalised to the island width at $\inpc{5}=1$. This is shown in figure \ref{fig:qfm_calculations} f-h) for the three W7-X scans. $\dqfm$ broadly reproduces the trends of $\dmin$ and $\dmax$ for edge $\iota=5/5$ and $5/6$; the island size increases at an accelerating rate as $\inpc{5}$ is reduced. For edge $\iota=5/4$, $\dqfm$ also reproduces the island phase transition; the changing sign of $A_{m,n}$ near $\inpc{5}=0.8$ causes the island size to vanish according to equation \eqref{eq:island_width}. The agreement between $\dqfm$ and $\dmin$ and $\dmax$ is worst for $\iota=5/6$.

\section{Novel HSX configurations}\label{sec:hsx_selected_configs}
\begin{figure}
    \centering
    \includegraphics[width=0.7\linewidth]{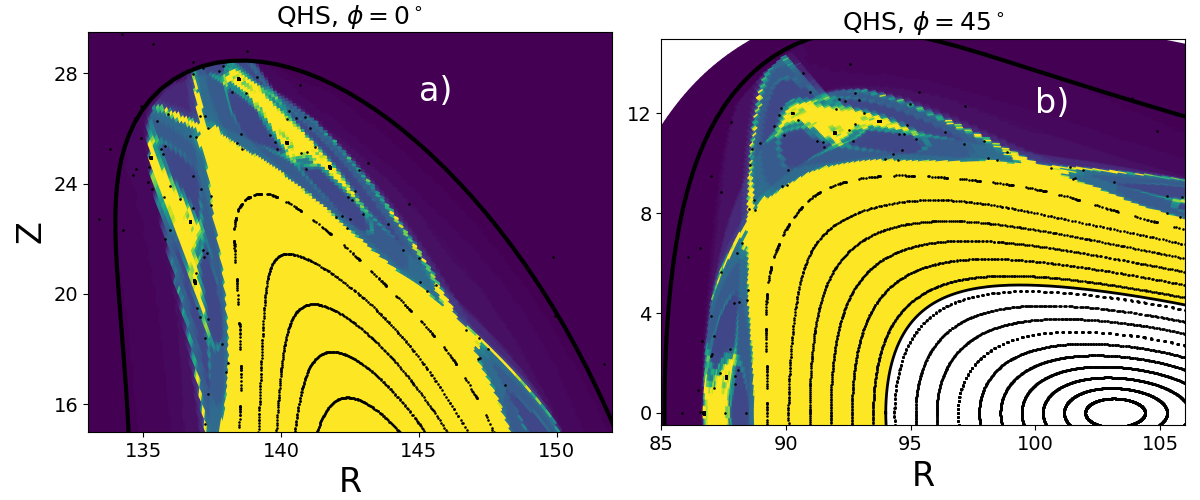}    
    \includegraphics[width=\linewidth]{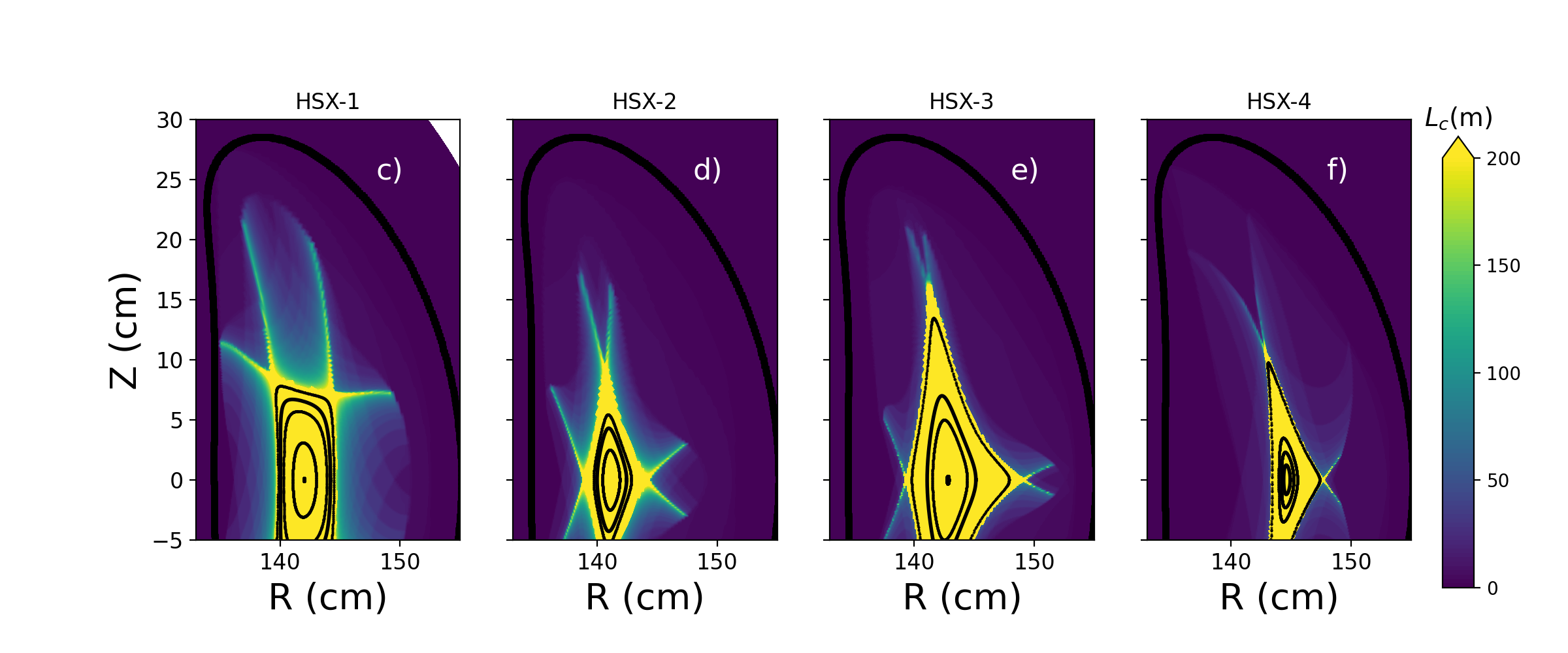}
    \includegraphics[width=0.7\linewidth]{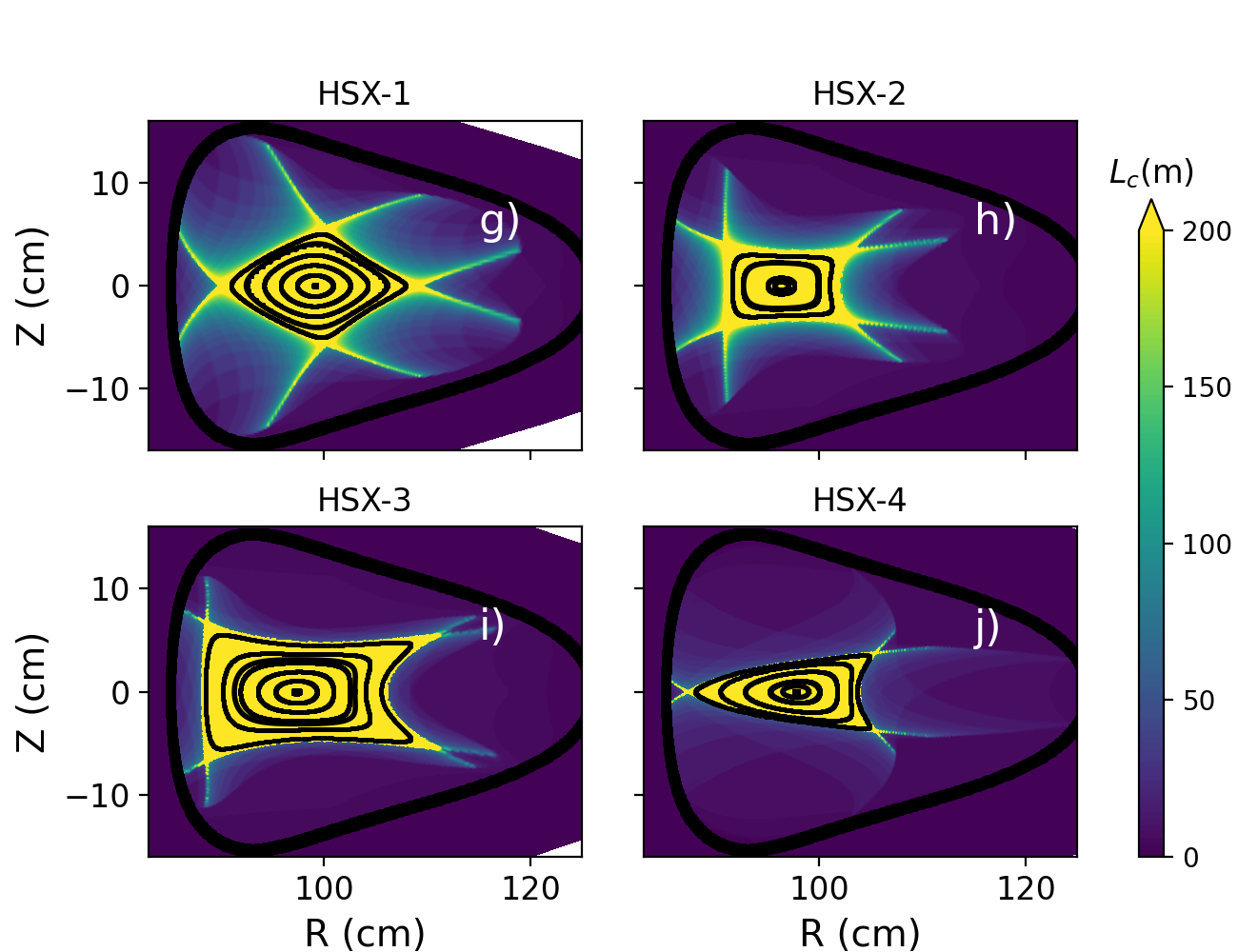}
    \caption{
    Wall-to-wall connection length $L_C$ for QHS and selected HSX configurations at $\phi=0^\circ$ and $\phi=45^\circ$, with Poincaré data shown in black. a-b) QHS configuration. c-f) Selected HSX configurations at $\phi=0^\circ$. g-j) Selected HSX configurations at $\phi=45^\circ$.
    }
    \label{fig:hsx_3d_illustrations}
\end{figure}
\begin{figure}
    \centering
    \includegraphics[width=0.8\linewidth]{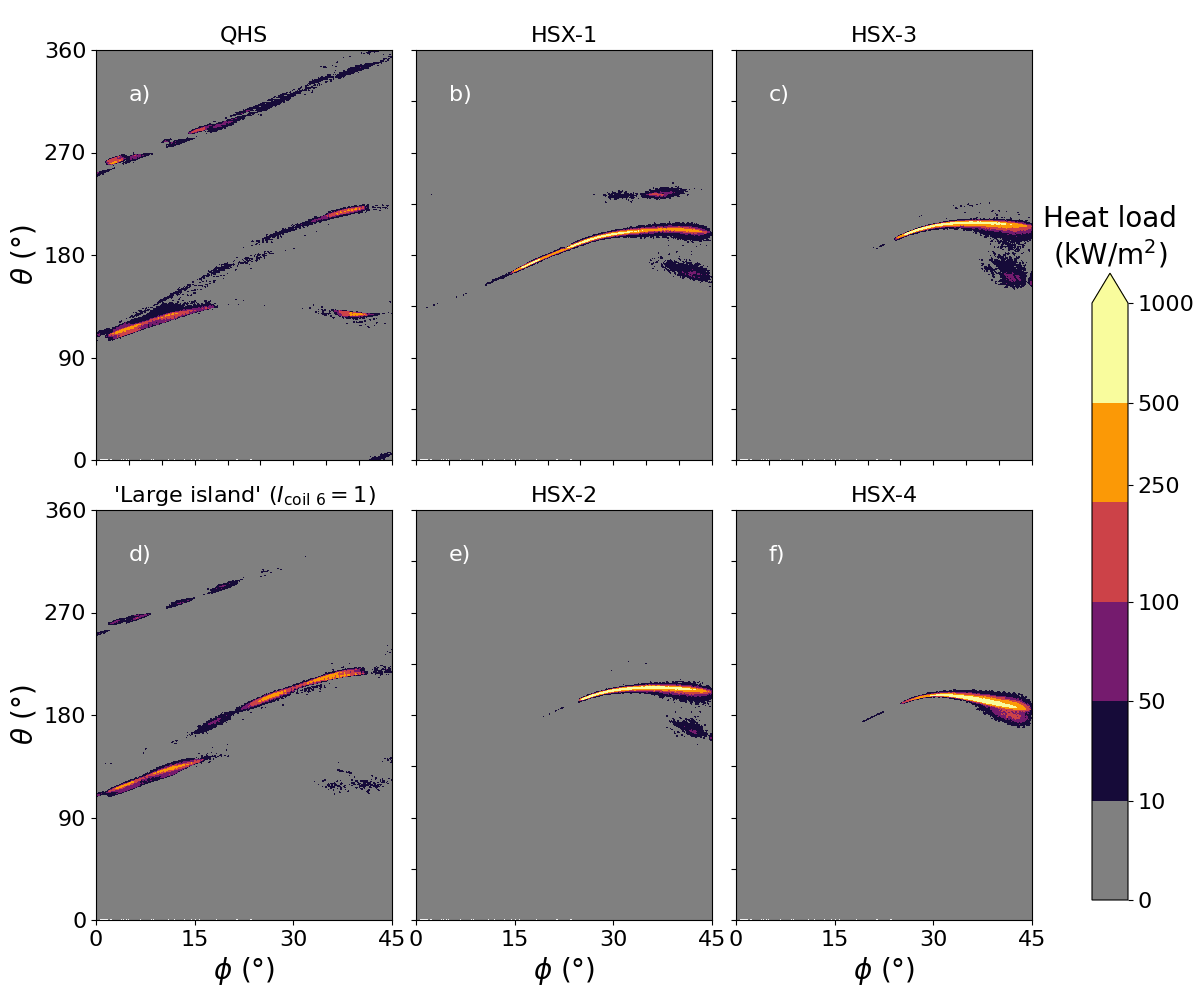}    
    \caption{
    Simulated heat loads $Q$ on the HSX vessel wall for reference (QHS, large island) and selected configurations for assumed non-radiated escaping power $\psol=20$kW, as a function of toroidal angle (over one half field period i.e. $\phi=0^\circ$ to $\phi=45^\circ$) and poloidal angle $\theta$, defined geometrically with respect to the QHS configuration magnetic axis. Grey region indicates where $Q$ is below a minimum cutoff value of $10\text{kW}/\text{m}^2$.
    }
    \label{fig:hsx_thetaphi_lc_and_depo}
\end{figure}
\begin{table}
\centering
\begin{tabular}{|c||c|c|c|} \hline \hline
Config. label & $\imain{6}$ & [$\iaux{1}$, $\iaux{2}$, $\iaux{3}$, $\iaux{4}$, $\iaux{5}$, $\iaux{6}$] & Edge $\iota$     \\ \hline 
QHS & $1$ & [$0, 0, 0, 0, 0, 0$] & $8/7$   \\  \hline
``Large island" & $1$ & [$0.1, 0.1, 0.1, 0.1, 0.1, 0.1$] & $4/4$   \\  \hline
HSX-1 & $0.4$ & [$0.1, 0.1, 0.1, 0.1, 0.1, 0.1$] & $4/4$   \\  \hline 
HSX-2 & $0.1$ & [$0.1, 0.1, 0.1, 0.1, 0.1, 0.1$] & $4/4$   \\  \hline 
HSX-3 & $0$ & [$-0.1, -0.1,  -0.1, 0.1, 0.13, 0.13$] & $4/4$   \\  \hline 
HSX-4 & $0$ & [$-0.2, -0.2, -0.2, -0.2, 0.029, 0.029$] & $4/3$   \\  \hline 
\end{tabular}
\caption{Coil winding currents for selected HSX configurations with giant edge structures. Currents normalised to order unity. To convert these currents to normalised currents through the filaments shown in figure 1, the auxiliary coil currents should be multiplied by 14.}
\label{tab:hsx_config_table}
\end{table}
We select four HSX configurations, with coil data shown in table \ref{tab:hsx_config_table}. The connection length and Poincaré data is shown in figure \ref{fig:hsx_3d_illustrations}, with the ``standard" QHS configuration also shown in \ref{fig:hsx_3d_illustrations} a-b) for reference. For the four selected configurations, $\imain{1-5}=1$. Configurations HSX-1,-2 and -3 have edge $\iota=4/4$, with $\imain{6}=0.4, 0.1, 0$ and are shown in figure \ref{fig:hsx_3d_illustrations} c-e), and g-i). Configuration HSX-4 has edge $\iota=4/3$ and $\imain{6}=0$ (figure \ref{fig:hsx_3d_illustrations} f) and j ). Setting $\imain{6}=0$ causes the confined plasma to shrink and inward-shift at $\phi=45^\circ$, such that the plasma starts to approach a limiter configuration. However, we partially alleviate this problem in HSX-3 and -4 by tuning $\iaux{1-6}$ to inward/outward shift the plasma. This is rather similar to the role of the planar coils in W7-X, and the mechanism by which these coils create inward/outward shift is described in Appendix \ref{app:shift_explained}.

For the four configurations, $L_C$ shows a lack of the spatially complex and fine structure seen in QHS; the magnetic structure which is visible to the plasma has little chaos. The X-point manifolds coincide with large $L_C$, since field lines on and near the manifolds move exponentially slowly in $(R,Z)$ in the neighbourhood of the X-point. 

The simulated plasma heat loads on the vessel wall is shown in figure \ref{fig:hsx_thetaphi_lc_and_depo}, plotted as a function of cylindrical toroidal coordinate $\phi$ and geometric poloidal angle $\theta=\arctan((Z-Z_\text{ma})/(R-R_\text{ma}))$ (where $R_\text{ma}(\phi)$ and $Z_\text{ma}(\phi)$ is the magnetic axis of the QHS configuration). These heat loads are calculated using the code EMC3-Lite \citep{feng2022review}, which uses an anisotropic heat diffusion model with spatially uniform conductivities. The parallel thermal diffusivity is set by the Spitzer conductivity, for which we use input parameters $T_e=T_i=50$eV, and the perpendicular diffusivity is set by the plasma density (for which we assume $n_e=10^{18}\text{m}^{-3}$) and a thermal diffusion coefficient which we take as $\chi=1\text{m}^2/\text{s}$. These values are selected based on a comparison of EMC3-Eirene simulations of HSX with Langmuir probe measurements \citep{akerson2016identification} (for further discussion see \cite{davies2024semi}). We assume a total non-radiated power exhaust $\psol=20$kW. For reference we also include plots for two reference configurations, the QHS configuration and edge $\iota=4/4$ $\imain{6}=1$ configuration (similar to the configuration discussed elsewhere and labelled ``large island" \citep{bader2013simulations, bader2017hsx, garcia2025_hsx}).

The simulated heat deposition shows notable differences between these configurations. QHS and large island configurations (figure \ref{fig:hsx_thetaphi_lc_and_depo} a) and d)) show the heat deposited in a helical strike pattern (i.e. diagonal lines in $(\phi, \theta)$). As $\imain{6}$ decreases the strike pattern becomes more toroidally localised around $\phi=45^\circ$ and poloidally localised around $\phi=180^\circ$ (the inboard side of the machine). This coincides both with where $L_C$ on the wall is greatest and the ``legs" formed by the island chain separatrix. Although there are in total 8 or 6 divertor legs (for edge $\iota=4/4$ or $4/3$ respectively), within a single half-field period most of these do not deposit heat on the wall; they have intersected the wall in a different half-field period and are therefore ``shadowed" in this half field period (which can be inferred by $L_C$ on the divertor leg becoming small). For example, figure \ref{fig:hsx_thetaphi_lc_and_depo} b) and c) show three and two distinct strike locations respectively, associated with two or three unshadowed divertor legs intersecting the wall in this half field period. 

For HSX-3 and HSX-4 (figures \ref{fig:hsx_thetaphi_lc_and_depo} c) and d)) the toroidal localisation increases further and the strike lines become wider poloidally, suggesting a partial transition towards limiter-like behaviour, with perpendicular transport from the confined region depositing heat on the wall, bypassing the parallel transport channel of the divertor legs. The toroidal concentration of plasma increases the peak power per unit area on the wall, which is unlikely to be problematic in HSX (due to the relatively short pulse length), but may be undesirable if extrapolated to a larger and higher-powered device. However, the relative geometric simplicity of the edge may facilitate tight baffling and geometric closure of the divertor, which could be beneficial for achieving a stable high-radiative regime and thus alleviating thermal limits on PFCs. Such tight baffling is challenging for the QHS configuration \citep{boeyaert2025towards}.

The neoclassical transport properties of these configurations are not targeted. Subsequent calculation of the neoclassical transport (characterised by the ``effective ripple" $\epsilon_\text{eff}$, calculated by the NEO code \citep{nemov1999evaluation}) finds the  neoclassical transport to be severely degraded in these configurations ($0.3 \leq \epsilon_\text{eff} \leq 2 $ across the minor radius, compared with $\mathcal{O}(0.01)$ for the standard and large island configurations), due to a degradation of helical symmetry. Running such experiments would therefore be comparable to non-optimised stellarators and would likely severely limit the core plasma temperature, but could nevertheless be informative for edge studies.

\section{W7-X: experimentally plausible selected configurations} \label{sec:w7x_selected_configs}
\begin{figure}
    \centering
    \includegraphics[width=0.98\linewidth]{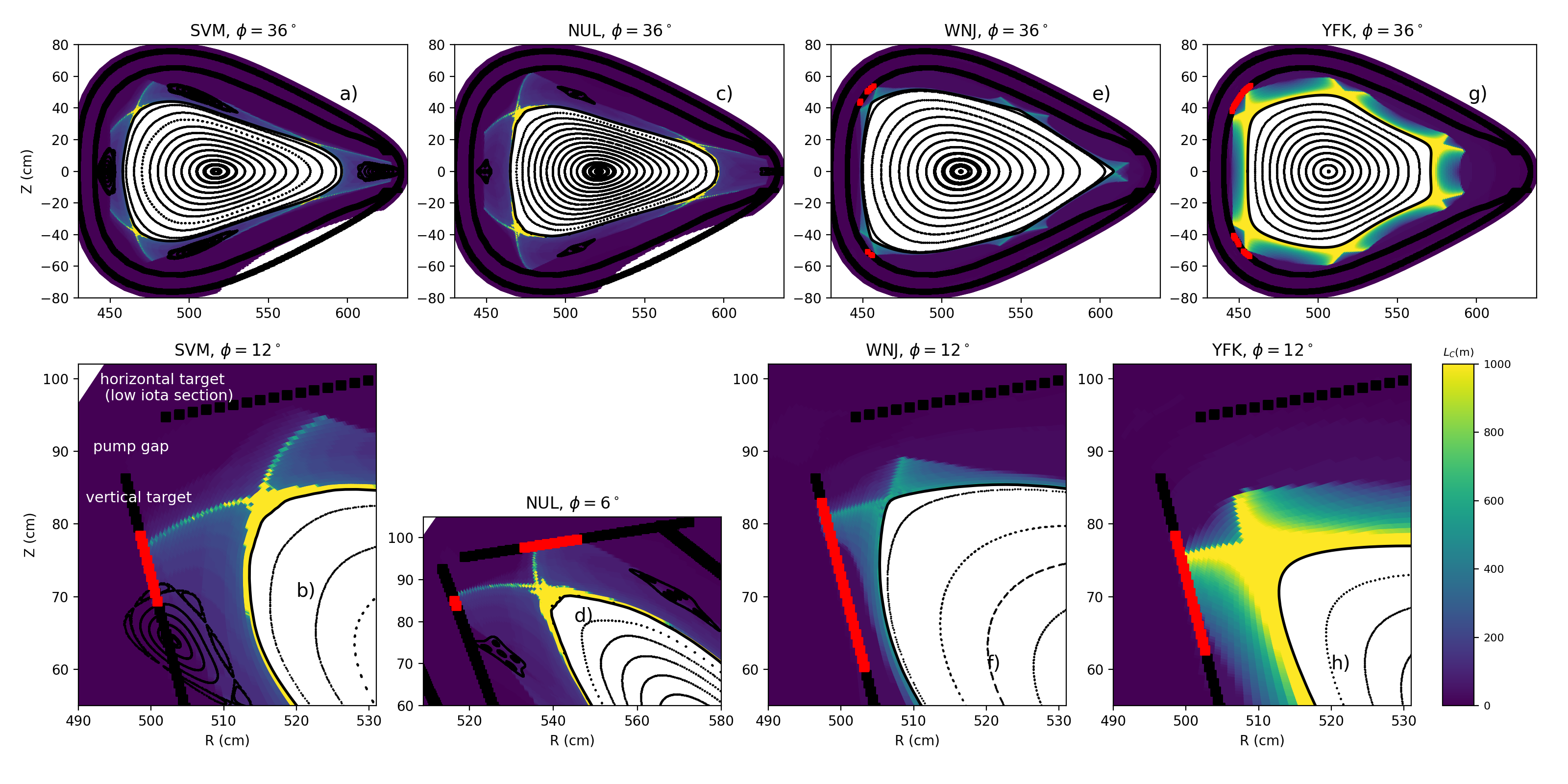}
    \caption{
    PFC-to-PFC connection length $L_C$ for candidate W7-X configurations at different toroidal locations. The locations on PFCs where the simulated heat loads exceed a minimum cutoff value of $1\text{MW}/\text{m}^2$ for non-radiated power $\psol=5$MW are shown in red. a-b) SVM configuration; c-d) NUL ; e-f) WNJ ; g-h) YFK.
    }
    \label{fig:w7_cool_configs}
\end{figure}
\begin{figure}
    \centering
    \includegraphics[width=0.8\linewidth]{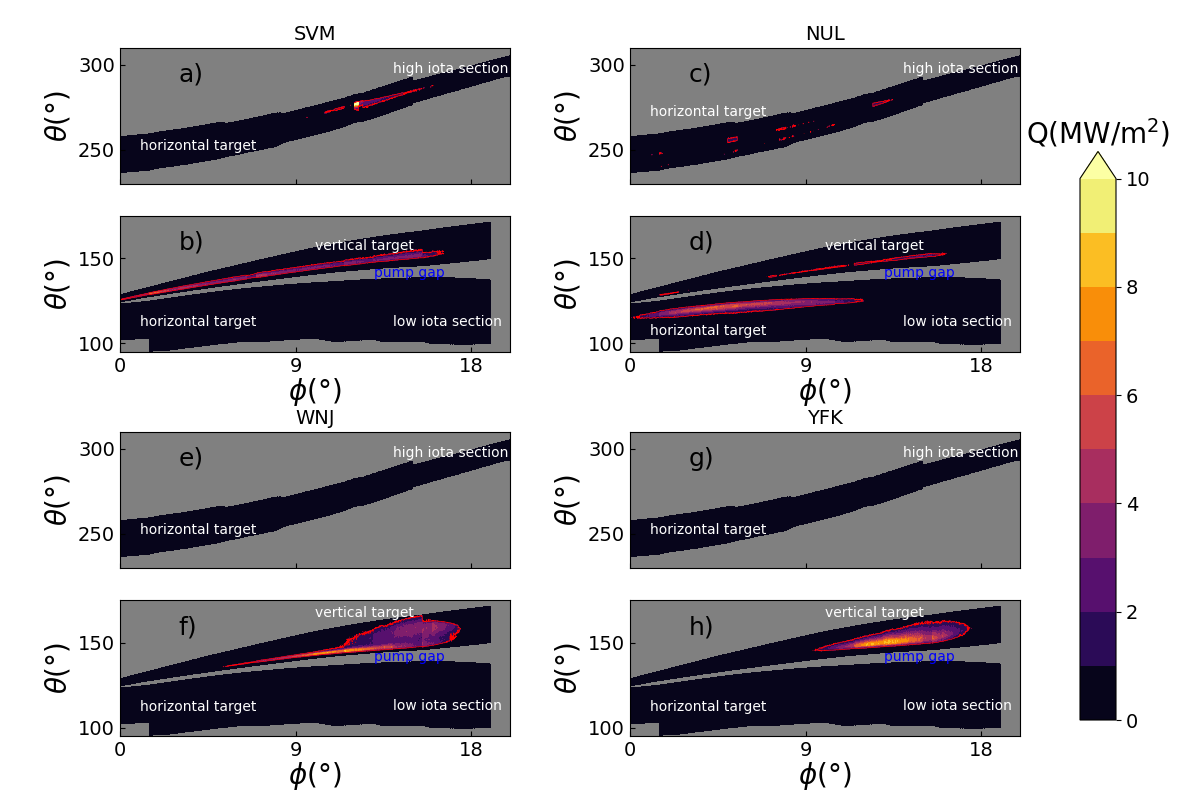}
    \caption{
    Heat loads on divertor plates for selected W7-X configurations as a function of toroidal angle $\phi$ (restricted to the region of interest) and geometric poloidal angle $\theta$ for assumed non-radiated power $\psol=5$MW. Red outline indicates where the heat load exceeds $1\text{MW}/\text{m}^2$. Heat loads on vessel wall and other plasma-facing components not shown. 
    }
    \label{fig:w7_cool_configs_heat_loads}
\end{figure}
\begin{figure}
    \centering
    \includegraphics[width=0.8\linewidth]{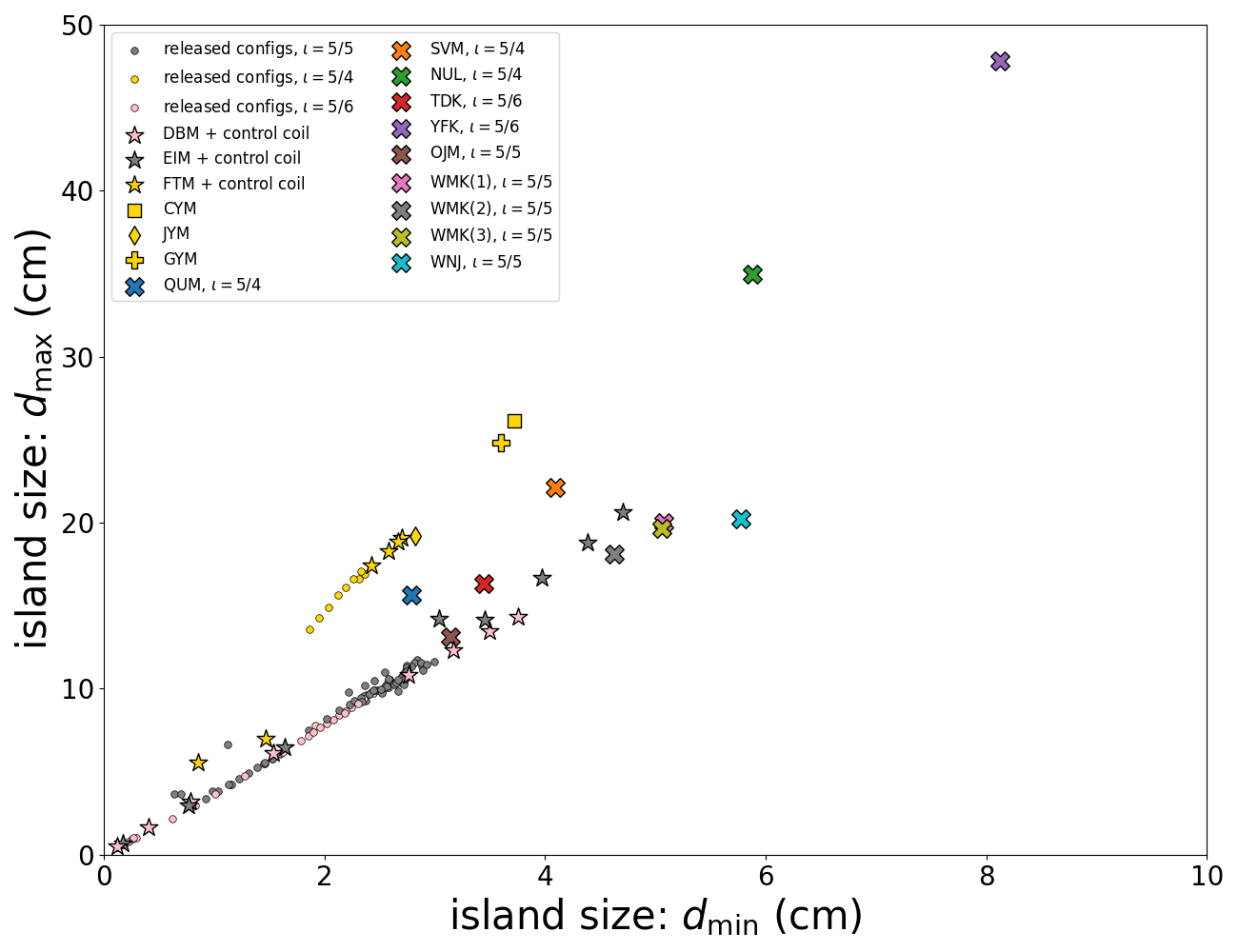}
    \caption{
    Island size for W7-X released and proposed configurations. All configurations are vacuum fields and have zero control coil current unless otherwise specified. This includes internal island chains and island chains which act as island divertors.
    }
    \label{fig:w7x_all_comparison}
\end{figure}
\begin{figure}
    \centering
    \includegraphics[width=0.6\linewidth]{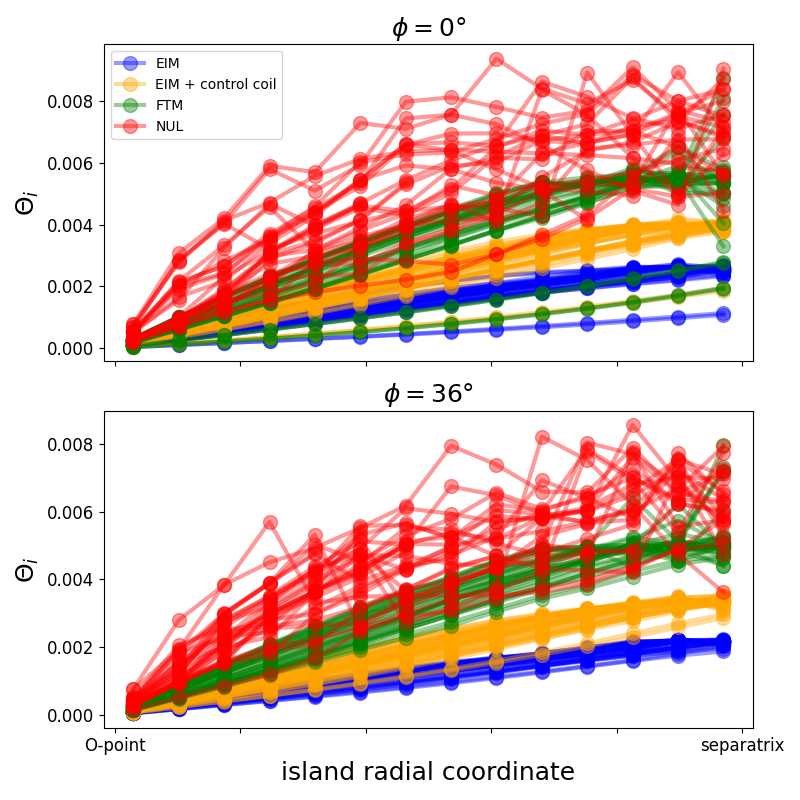}
    \caption{
    Field line pitch calculation for selected field lines within the islands for selected W7-X configurations. Upper: field lines sampled at $\phi=0^\circ$. Lower: field lines sampled at $\phi=36^\circ$.}
    \label{fig:w7x_field_line_pitch}
\end{figure}
\subsection{Experimental considerations}
Reducing $\inpc{5}$ (as shown in figure \ref{fig:hsx_w7x_1d_scans_poincare}) creates far-extending divertor legs which carries the plasma to the PFCs. However, similarly to HSX, the strike pattern becomes strongly localised around the toroidal location of the manipulated coil (since the toroidal magnetic flux is conserved but the field strength drops around $\phi=36^\circ$), meaning that the PFC heat loads are not necessarily compatible with engineering limits. To promote experimental feasibility, we adjust the planar coil currents to raise/lower $\iota$ (thus pushing the resonance inwards/outwards with respect to the magnetic axis) and shift the configuration inwards/outwards in major radius. This process is performed manually to create candidate configurations.

We select specific configurations based on several considerations: (1) high power fraction caught on divertor plates; (2) low peak loads (power per unit area) on walls, baffles and shields; (3) maximising island size; (4) minimising the coil current deviation from the normal operating range of W7-X; (5) minimisation of chaos for large islands. Of these, (3) and (4) are mostly obviously in conflict. The reason for considering (5) is because large islands (for example, using control coils) usually results in chaos, which have multifarious and little-understood consequences. Experiments which combine large islands with low chaos would allow these effects to be studied independently. 

As with HSX, the neoclassical transport properties are not targeted and are subsequently found to be poor ($\epsilon_\text{eff} \sim 0.1$). It is worth noting that configurations with $\epsilon_\text{eff} \sim 0.08$ have run experimentally in W7-X, but had limited core temperatures of $T_e \leq 1$keV.

\subsection{Selected configurations}
\begin{table}
\centering
\begin{tabular}{|c||c|c|c|} \hline \hline
Name & [$\inpc{1}, \inpc{2}, \inpc{3}, \inpc{4}, \inpc{5}$] & [$\ipca, \ipcb$] & Edge $\iota$     \\ \hline 
``standard" (EIM) & [$1,1,1,1,1$] & [$0, 0$] & $5/5$   \\  \hline 
``low iota" (DBM) & [$1,1,1,1,1$] & [$0.25, 0.25$] & $5/6$   \\  \hline 
``high iota" (FTM) & [$1,1,1,1,1$] & [$-0.23, -0.23$] & $5/4$   \\  \hline 
SVM & [$1,1,1,1,0.6$] & [$-0.23, -0.23$] & $5/4$   \\  \hline 
NUL & [$0.7,1,1,1,0.5$] & [$-0.23, -0.19$] & $5/4$   \\  \hline 
WNJ & [$1,1,0.5,1,0.23$] & [$-0.08, 0.08$] & $5/5$   \\  \hline 
YFK & [$1,1,0.5,1,0.1$] & [$0.13, 0.25$] & $5/6$   \\  \hline 
\end{tabular}
\caption{Coil filament currents for the paradigmatic ``standard", "``low iota" and ``high iota" configurations and selected W7-X configurations with giant edge structures. Currents are filament currents (i.e. currents in the filaments shown in figure \ref{fig:hsx_3d_illustrations}), normalised to order unity.}
\label{tab:w7x_config_table}
\end{table}
Our configuration search yields 10 possibly interesting configurations, and we present analysis for four of these, which in the W7-X nomenclature are named: SVM, NUL, WNJ and YFK. This three-letter code bins configurations according to mirror ratio (the first letter, with an alphabetically earlier letter indicating lower mirror ratio), on-axis $\iota$ (the second letter, with an earlier letter indicating lower $\iota$) and inward/outward shift (the third letter, earlier indicating outward-shifted). The coil currents for all named W7-X configurations are given in table \ref{tab:w7x_config_table}. Poincaré and $L_C$ information is shown in figure \ref{fig:w7_cool_configs} and heat loads on divertor plates are shown in figure \ref{fig:w7_cool_configs_heat_loads} for input parameters ($T=100$eV, $n_e=10^{19}\text{m}^{-3}$, $\chi=3\text{m}^2/\text{s}$, $\psol=5$MW). These $T$, $n_e$, and $\chi$ values are the values found by Gao \textit{et al} to best match experimental data in attached conditions \citep{gao2023improvement}. Similar EMC3-Lite modelling work has used comparable values \citep{gao2024heat, naujoks2026heat}, although an EMC3-Eirene study (which uses a different physics model) found a better match with experimental data with lower $\chi$ values \citep{bold2022parametrisation}.

Two of these configurations, SVM and NUL, have edge $\iota=5/4$, with $\inpc{5}=0.6$ and $0.5$ respectively. SVM comes directly from the scan presented in section \ref{sec:1d_scans}. The NUL configuration uses $\ipca$ and $\ipcb$ to outward shift the configuration and reduces $\inpc{1}$ to ensure the heat falls mostly on the divertor plates. As can be seen in figures \ref{fig:w7_cool_configs} a-d), the island separatrix forms a leg of high $L_C$ which transports heat onto the PFCs; figure \ref{fig:w7_cool_configs_heat_loads} a-d) shows that this creates toroidally long strike lines on the divertor plates. 

Both the WNJ configuration (edge $\iota=5/5$) and YFK (edge $\iota=5/6$) are outward-shifted configurations and reduce $\inpc{3}$ to control the PFC heat load distribution. Figures \ref{fig:w7_cool_configs} e-f) show that WNJ has limiter-like features with the divertor plate coming very close to the confined region and resulting in a poloidally broad heat deposition pattern in figure \ref{fig:w7_cool_configs_heat_loads} f). The divertor plate comes less close in the YFK configuration (figures \ref{fig:w7_cool_configs} g-h)), but the divertor plate heat load remains broad (figure \ref{fig:w7_cool_configs_heat_loads} h)), presumably due to the large $L_C$ in the edge, which causes perpendicular transport to play a larger role and spread the heat across flux surfaces to a greater extent.

In the four selected configurations, the majority of the heat lands on the divertor plates according to EMC3-Lite simulations (82\%, 84\%, 94\%, and 77\% for SVM, NUL, WNJ and YFK respectively) for the input parameters selected, with the remainder of the heat caught on the baffles or heat shields. Whether these configurations can be safely run in experiment (from the perspective of both heat loads and coil forces) and whether more-feasible nearby configurations can be found is a topic of ongoing research. The maximum acceptable steady-state heat loads on the heat shields are $250$kw/m \citep{peng2013thermo}, and therefore running configurations with low $\psol$ is necessary, particularly for WNJ and YFK. 

\subsection{Island size quantification}
It is instructive to compare the island size (characterised by $\dmin$ and $\dmax$) for these configurations with the released configurations from the of W7-X database (i.e. all the vacuum magnetic configurations which are approved to run experimentally). To do this, we take all approved configurations (each of which corresponds to a set of values for $\inpc{1-5}$ and $I_\text{PC A,B}$), normalise the coil filament currents so that $\inpc{1}=1$, and then remove duplicate configurations, defined as configurations where the maximum deviation in coil filament currents from any non-duplicated configuration is below $1\%$ (this occurs, for example, when the same configuration is run at a different field strength.) For each non-duplicated configuration, we perform the $\dmin$ and $\dmax$ calculation presented in section \ref{sec:island_size_quantification_d}, assuming $\icc=0$. We include the calculation for internal island chains and island divertor island chains, but not for island chains which are invisible to the plasma (i.e. the inner island separatrix intersects or is entirely beyond the PFCs). In addition to the released configurations, for the three paradigmatic configurations ``standard", ``high iota", ``low iota", we scan $\icc$ over the range $\icc=\pm 40/1402$, which corresponds to $\pm 40$kA filament current ($5$kA winding current $\times 8$ windings) at normal field strength (the normal maximum for long-pulse operation is 20kA filament current). 

The results of this analysis is shown in figure \ref{fig:w7x_all_comparison}. The released configurations tend to fall into two branches; the branch of $\iota=5/5$ or $5/6$ and the branch of $\iota=5/4$. For both, $\dmin \propto \dmax$, but the $\iota=5/4$ has a steeper gradient (i.e. a larger $\dmax$ for a given $\dmin$.) Three configurations are shown explicitly; CYM \citep{boeyaert2025analysis}, JYM and GYM \citep{smiet2025turnstile}, which are part of a family of ``super-high-iota" configurations for which the relative planar coil currents are large and negative (as a consequence, the confined plasma is shrunk and the islands pushed inwards). CYM and GYM have notably large island size. It is also evident that island size can be increased via $\icc$, particularly for $\iota=5/5$ islands. However, in all of these cases $\dmin<5\text{cm}$ and $\dmax<30\text{cm}$. Ignoring the CYM, GYM and JYM configurations and the $\icc$ configurations, $\dmin<4\text{cm}$ and $\dmax<20\text{cm}$ for the released configurations.

Our selected configurations have large islands with respect to the normal operational space of W7-X and fall outside the normal branches of $\dmin$ and $\dmax$. Of all configurations, YFK has the largest $\dmin$ and $\dmax$ ($\sim 8$cm and $\sim 50$cm respectively), approximately double the maximum of released configurations. It is interesting to note from figure \ref{fig:w7_cool_configs} that it has the largest typical $L_C$ in the scrape-off layer. This is consistent with a low internal island pitch $\Theta_i$, suggesting low magnetic shear (rather than a large resonant perturbation) across the island. This is consistent with the analysis of section \ref{sec:island_size_quantification_qfm}.

\subsection{Field line pitch calculation}
To study the effect of giant islands on the internal island pitch $\Theta_i$, we calculate $\Theta_i$ for the typical W7-X configurations and NUL. $\Theta_i$ is defined by 
\begin{equation}
    \Theta_i = \frac{d_\perp}{d_\parallel},
\end{equation}
where $d_{\parallel}$ is the parallel (i.e. along the magnetic field $\BBB$) distance travelled by a field line to traverse $m$ field periods (where $m$ is set by the island resonance, $\iota=m/n$) and $d_\perp$ is the perpendicular distance travelled by the field line in this toroidal location after $m$ field periods i.e. $d_\perp = \sqrt{(R_{m} - R_0)^2 + (Z_m - Z_0)^2}$ where $(R_m, Z_m)$ is the location of a field line started at $(R_0, Z_0)$ and followed for $m$ field periods. $\Theta_i$ varies with the toroidal and poloidal location of the island, is small near the island O-point and vanishes on the O- and X-points. To find representative values of $\Theta_i$ we therefore sample multiple magnetic field lines within the islands, and perform the calculation for all islands at toroidal locations of $\phi=0^\circ$ and $\phi=36^\circ$. 

$\Theta_i$ is shown in figure \ref{fig:w7x_field_line_pitch} for configurations EIM, FTM and NUL, as well as ``EIM $+$ control coil", in which the control coils are used to make the islands large, and increases $\Theta_i$ accordingly. The NUL and FTM configurations generally have larger $\Theta_i$, which is expected since $d_\parallel$ is reduced by approximately $20\%$ ($m$ reduces from $5$ for EIM to $4$ for FTM/NUL). In addition, $\Theta_i$ generally increases with island size, with NUL having the largest average $\Theta_i$, reaching $0.006$ near the separatrix. These relatively large $\Theta_i$ values imply favourable $n_d/n_u$ scaling according to simple 2 point-like models.

\subsection{Magnetic chaos on the inner separatrix}
\begin{figure}
    \centering
    \includegraphics[width=0.8\linewidth]{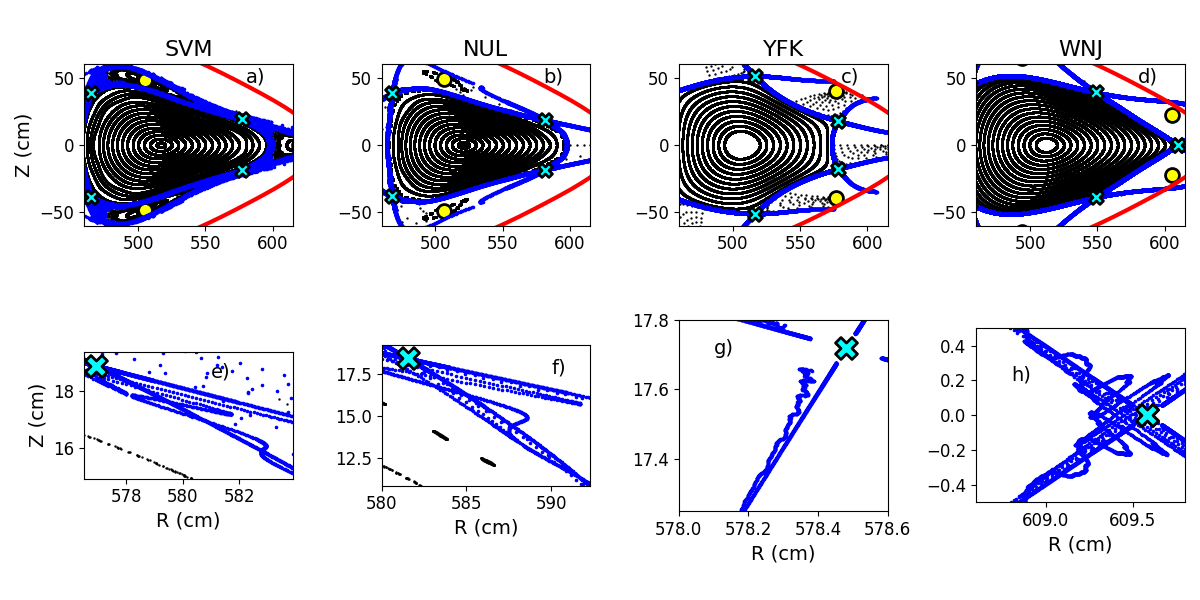}
    \caption{
    Poincaré for the 4 selected configurations at $\phi=36^\circ$. e-h) Zoom-in of manifolds near the X-points to show turnstile lobe structures on the inner separatrix.
    }
    \label{fig:w7x_selected_configs_chaos}
\end{figure}
Previous ``large island" W7-X configurations (such as the GYM, JYM and CYM configurations, and configurations with large $\icc$) have shown chaos in the edge magnetic field \citep{smiet2025turnstile, boeyaert2025analysis}. One might therefore expect significant signatures of magnetic chaos in these configurations. However, this is not the case, as shown by the $L_C$ plots in figure \ref{fig:w7_cool_configs}, and the Poincaré plots in figure \ref{fig:w7x_selected_configs_chaos}. As in the HSX examples, $L_C$ shows long, straight divertor legs in the edge, without large geometrically complex lobe structures in the island region or intruding into the main plasma. Figure \ref{fig:w7x_selected_configs_chaos} shows the inner boundary of the island, found by tracing the X-point manifolds. Whilst there are turnstile lobes, these are very small (the lobe area at $\phi=36^\circ$ are at most of the order $1\text{mm}^2$). It might be the case that the magnetic field in the edge (e.g. associated with the manifolds on the outside of the islands) is chaotic in some of these configurations, but this is ``invisible" to the plasma because of the locations of the plasma-facing components (field lines only sample the chaotic magnetic field for a small distance before intersecting PFCs.)

\section{Conclusions}\label{sec:conclusions}
This work presents simulations of magnetic configurations with giant edge structures in two existing modular coil stellarators, W7-X and HSX. In both cases, islands can be grown far beyond their typically reported range, demonstrating the flexibility of these machines for edge studies without needing to change coil geometry. Our simple method for increasing the island size significantly affects the neoclassical transport of these configurations. However, whilst undesirable for core performance, these could be useful edge experiments. In addition, they provide realistic magnetic fields which could be used by high-fidelity edge codes such as EMC3-Eirene to investigate the dependence of plasma properties (for example, downstream/upstream plasma density) on island size.

In this work we focus on a single coil in both devices which acts as a control knob for island size, that nearest the symmetric plane in the straight section. This appears to be a sufficient (but not necessary) condition for creating giant edge structures. We characterise the island width in three ways; $\dmin$, $\dmax$ (which are measures of distance of the island O-points to the inner separatrix) and an island width calculation calculated by modelling the island as a (spectrally pure) resonant perturbation to a hypothetical set of nested flux surfaces. The latter method is implemented by using \texttt{simsopt} to construct Quadratic Flux-Minimising surfaces, and has the advantage that both the shear of the unperturbed system and the resonant perturbation can be calculated. 

For HSX we use the auxiliary coils to select either edge $\iota=4/3$ or edge $\iota=4/4$. We identify four notable configurations where $\imain{6}$ is reduced to create giant edge structures, with long divertor legs intersecting the vessel wall. We use differential auxiliary coil currents to inward/outward shift the plasma, which increases the confined volume without limiting on the vessel wall. 

For W7-X, we identify several configurations with very large islands with edge $\iota=(5/4, 5/5, 5/6)$, in which the simulated PFC heat loads (using an anisotropic heat diffusion model, EMC3-Lite) fall mostly on the divertor plates. These islands are far beyond the existing configuration space, although the coil forces must be carefully considered before these experiments can be run. For one configuration, labelled NUL, the field line pitch in the island $\Theta_i$ is calculated at various points inside the island and is found to be generally larger than the typical ``standard" and ``high iota" configurations. This is advantageous for the downstream/upstream plasma density ratio ($n_d/n_u)$ according to simple models.

Beyond the experimental prospects in HSX and W7-X, these studies provide a examples for exploring novel edge structures in low-shear modular coil stellarators. Open questions which could be answered by existing edge codes is whether the scaling plasma density and temperature on the divertor legs resembles the tokamak 2-point model or the island divertor variations which have been proposed for stellarators. 

If these giant edge islands are desirable for plasma transport and/or geometric simplicity in divertor plate/baffle design, we speculate how unpaired X-points might be deliberately engineered when designing future modular-coil stellarators. The common feature between HSX and W7-X is a large toroidal extent around the straight section of the stellarator ($\phi=45^\circ$/$\phi=36^\circ$ for HSX/W7-X) of relatively low coil current. It might be that this is a sufficient (or at least, helpful) ingredient for making extremely large edge structures, although further research would be required to verify this. If so, coil optimisation could promote these edge features by optimising coil placement to leave a suitably toroidally large current gap. To conclude, the causes and consequences of giant islands and unpaired X-points are relatively poorly understood but are interesting in the context of stellarator divertor optimisation, and this work advances a means for their study, both computationally and perhaps experimentally.

\section{Acknowledgements}
This work has been carried out within the framework of the EUROfusion Consortium, partially funded by the European Union via the Euratom Research and Training Programme (Grant Agreement No 101052200 — EUROfusion). Views and opinions expressed are however those of the author(s) only and do not necessarily reflect those of the European Union or the European Commission. Neither the European Union nor the European Commission can be held responsible for them. This work was funded by the U.S. Department of Energy under grant numbers DE-FG02-93ER54222 and DE-SC0014210.

The authors would like to thank Michael Gerard, Eric Andrews and Eduardo Rodriguez for their valuable contributions and discussions.

\section{Declaration of Interests}
The authors report no conflict of interest.

\section{Data Availability}\label{sec:data_availability}
The data used in this study is available upon reasonable request. The tools used to generate the data presented in section \ref{sec:island_size_quantification_qfm} are available at \url{https://github.com/louisday/qfm}.

\bibliographystyle{jpp}
\bibliography{references}

\begin{appendix}

\section{Coil current scans in HSX}\label{app:hsx_coil_scans}
\begin{figure}
    \centering
    \includegraphics[width=0.95\linewidth]{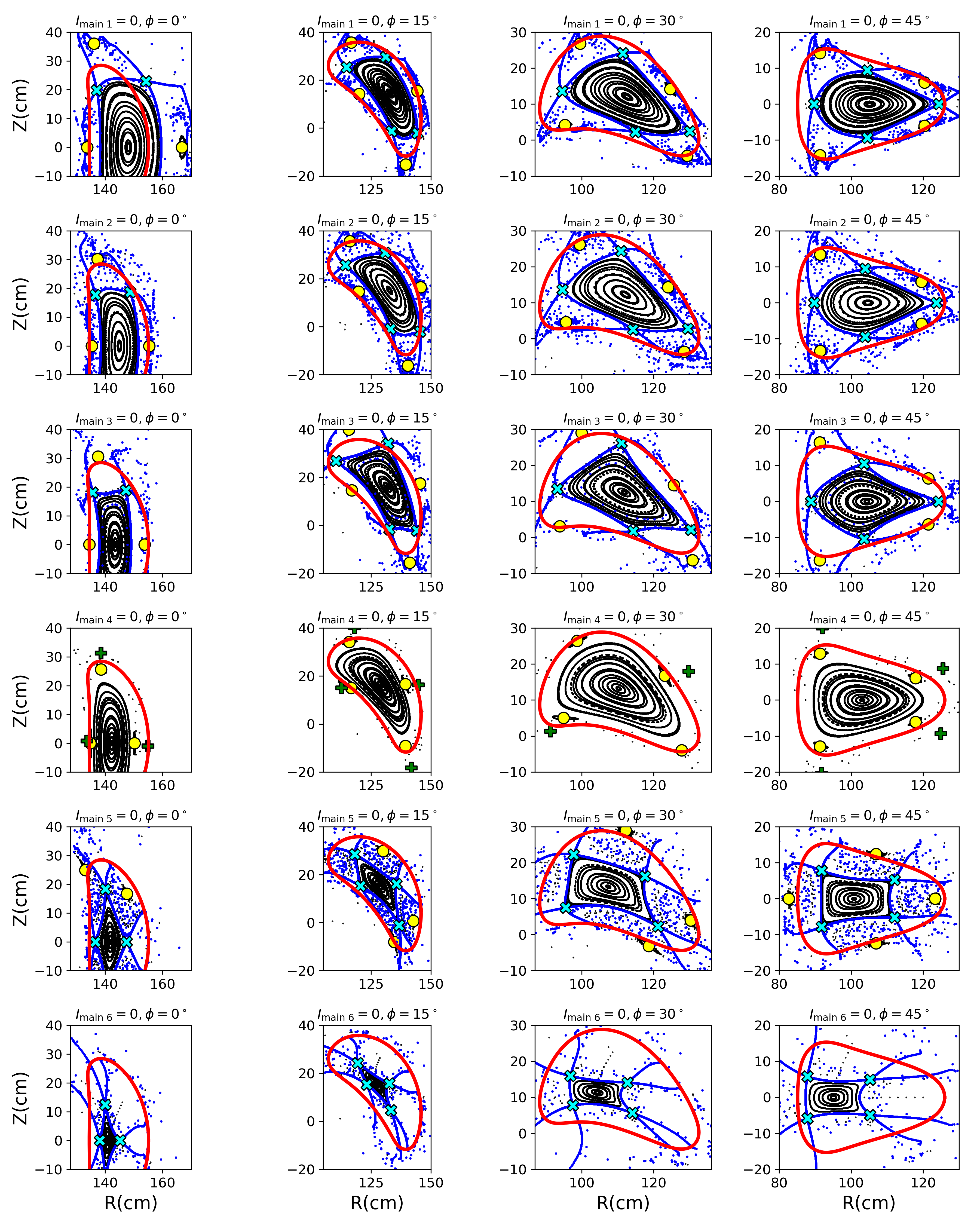}
    \caption{
    Magnetic structure in HSX, as each main coil is set to zero in turn. $\iaux{1-6}=0.1$ to ensure edge $\iota=4/4$. The lowest row corresponds to configuration HSX-3 in section \ref{sec:hsx_selected_configs}.
    }
    \label{fig:hsx_other_coils_scan}
\end{figure}
In light of the observation that adjusting the current in main coil 6 increases island size, it is worth elucidating whether changing other coil currents have a similar effect. A summary of this investigation is presented here, with the overall conclusion that modifying other coil currents can have a strong effect on island but have a tendency to create limiter-like configurations by moving the confined plasma to intersect the vessel wall.

The effect of setting to zero $\imain{1-6}$ is shown in figure \ref{fig:hsx_other_coils_scan}; in all cases we set $\iaux{1-6}=0.1$ to ensure edge $\iota=4/4$. In all cases, an $\iota=4/4$ edge structure is present. It can be seen that several coils are capable of producing large islands in the edge, but with $\imain{5}=0$ and $\imain{6}=0$ the most effective for creating giant edge structures with relatively long X-point legs. Setting $\imain{1-6}=0$ also changes the size and inward/outward shift of the plasma, with for example $\imain{1}=0$ creating a configuration which is grown and outward-shifted around $\phi=0^\circ$, limiting on the vessel wall. Similarly, $\imain{2-6}=0$ configurations are limiter- or almost-limiter-like, with the least limiter-like being $\imain{5}=0$ and $\imain{6}=0$. 

It might be that the tuning of $\iota$ and shift via $\iaux{1-6}$ might be able to create attractive experimental candidates for giant island configurations with $\imain{1-5}=0$ (with the most likely candidate being $\imain{5}=0$ since it is least limiter-like and displays long straight X-point legs), but these studies are left as future work. 

\section{Coil current scans in W7-X}\label{app:w7x_coil_scans}
\begin{figure}
    \centering
    \includegraphics[width=0.95\linewidth]{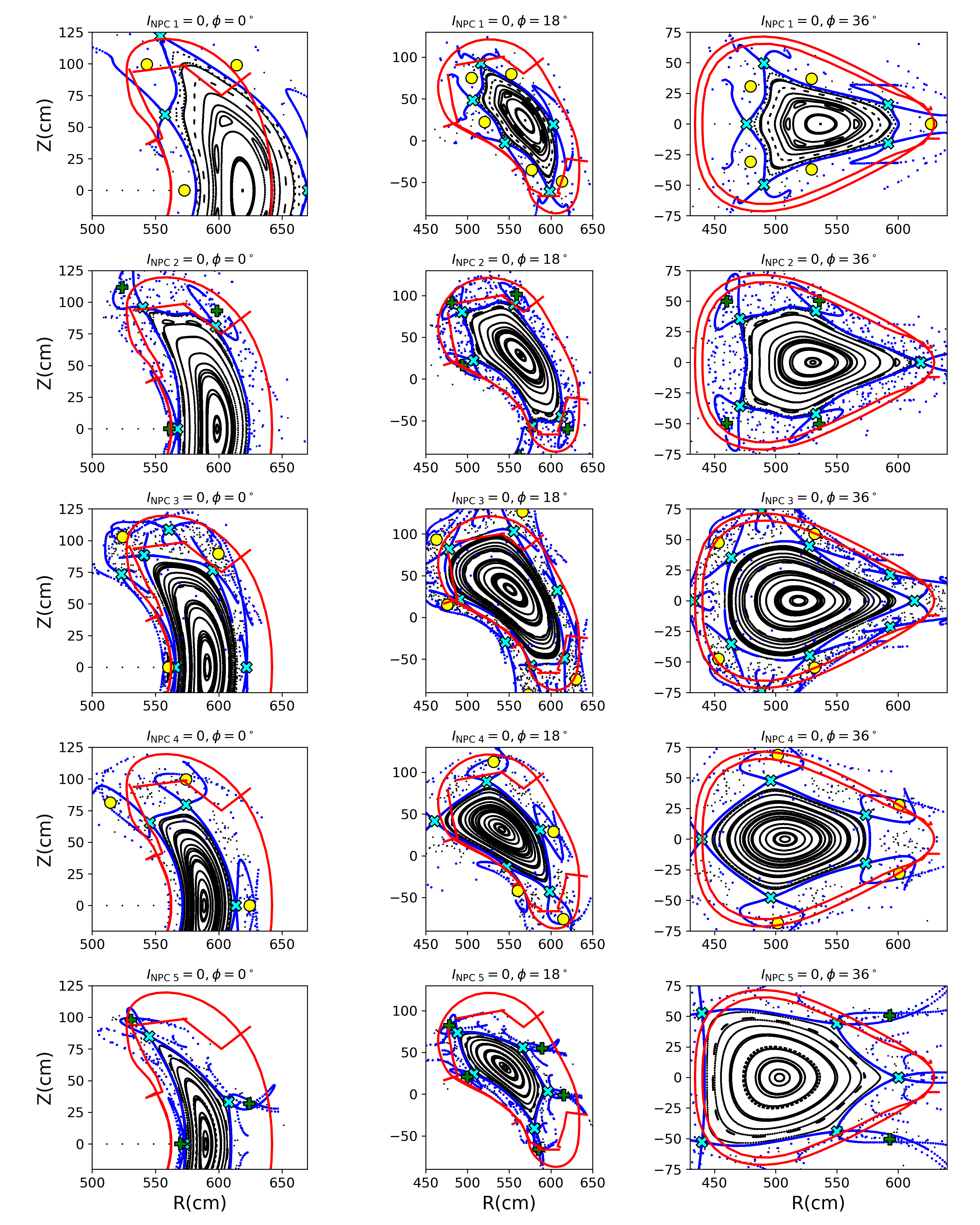}
    \caption{
    Magnetic structure in W7-X, as each non-planar coil is set to zero in turn.  $\ipca=\ipcb=\icc=0$ in all cases.
    }
    \label{fig:w7x_other_coils_scan}
\end{figure}
Similar to the analysis performed in appendix \ref{app:hsx_coil_scans}, it can be seen from figure \ref{fig:w7x_other_coils_scan} that strongly reducing the currents in non-planar coils 1-4 currents in W7-X. As in HSX, we observe growing and outward shifting of the plasma when $\inpc{1}=0$. In all cases, there is limiter-like behaviour. This is exacerbated in W7-X because unlike HSX it also contains divertor and baffle structures designed with a particular edge structure in mind. Preliminary investigations to find configurations in which $\inpc{1-4}$ are low and the simulated heat loads fall mostly on the divertor plates have thus far been unsuccessful. 

\section{Inward/outward shift in W7-X and HSX}\label{app:shift_explained}
\begin{figure}
    \centering
    \includegraphics[width=0.55\linewidth]{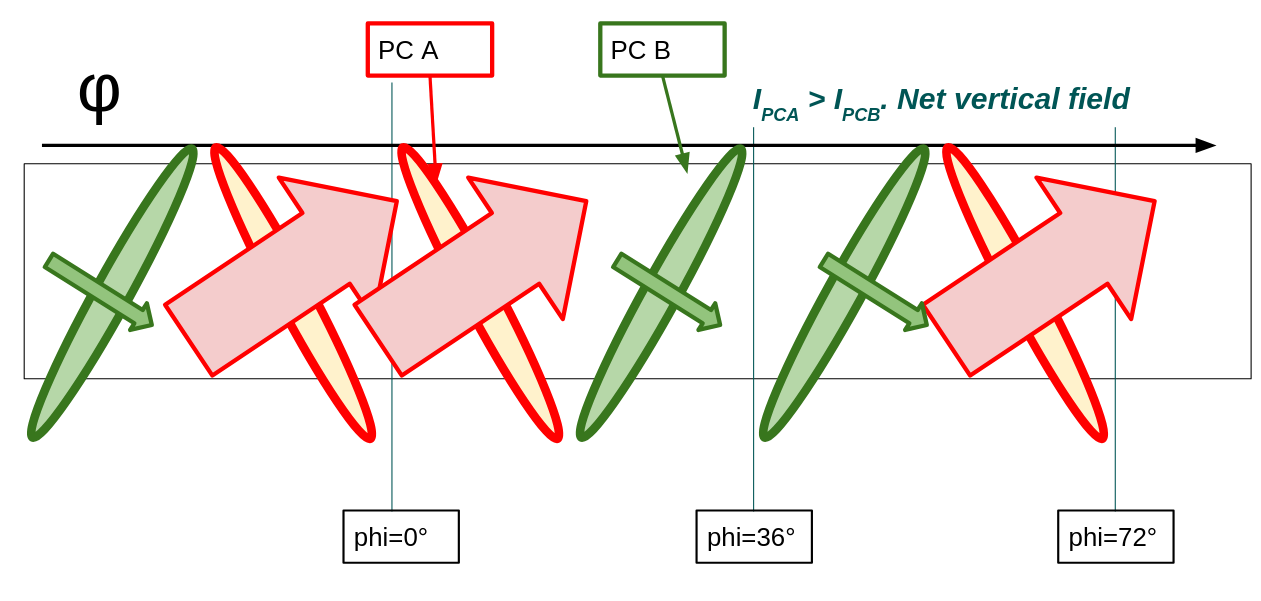}
    \includegraphics[width=0.4\linewidth]{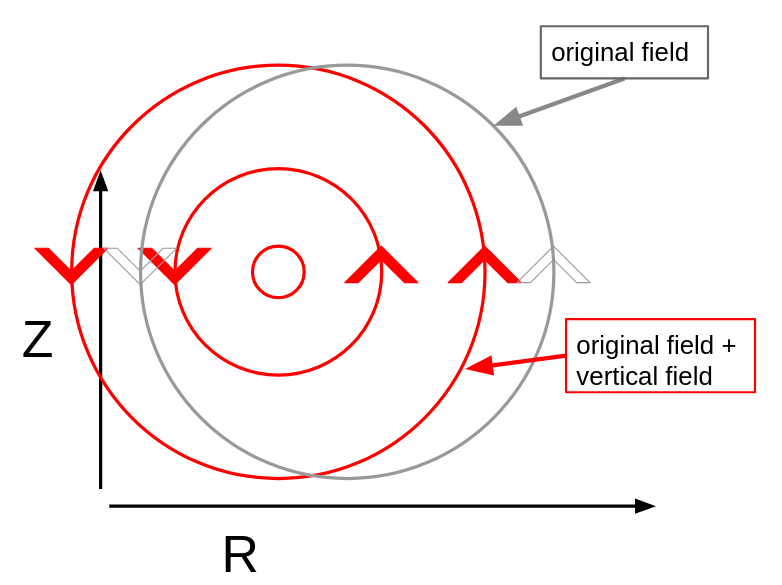}    
    \caption{
    Illustration of inward/outward shift in W7-X via planar coils A and B. Left: illustration of how tilted planar coils with differing currents can create a net vertical field. Right: illustration of how a vertical field can create an inward/outward shift in $R$.
    }
    \label{fig:shift_cartoon}
\end{figure}
We make the claim in section \ref{sec:hsx_selected_configs} that HSX's auxiliary coils provide inward/outward shift capabilities similar to W7-X, we present here a simple physics explanation to justify this claim.

The role of the coils on the inward/outward shift is illustrated in figure \ref{fig:shift_cartoon}, using W7-X planar coils A and B as an illustrative example. Because these coils are tilted (see figure \ref{fig:hsx_w7x_coils_3d}), they produce both a toroidal and vertical magnetic field, with planar coil A producing an upwards vertical field and planar coil B producing a downwards vertical field for the normal direction of current flow in the coils. When $\ipca \neq \ipcb$, the result is a net vertical magnetic field in addition to the toroidal field (the toroidal field goes like the sum of these currents, which explains why this boosts/reduces $\iota$). A net vertical field shifts the plasma inwards/outwards, as sketched in figure \ref{fig:shift_cartoon} b). This is most easily seen for an axisymmetric magnetic field, whereby the magnetic axis is defined by the location of $B_Z=0$, and therefore shifts in $R$ when an externally imposed $B_Z$ is applied.

By comparing  figure \ref{fig:hsx_w7x_coils_3d} c) and d), it can be seen than HSX auxiliary coils 1 and 6 loosely resemble W7-X coils A and B (circular coils tilted with respect to the vertical, so that viewed from the side they resemble a  ``\textbackslash" and ``/" respectively), with auxiliary coils 2-5 smoothly transitioning between the two. As a result, differential currents in the auxiliary coils can create a net vertical field, and consequently an inward/outward shift. 
\end{appendix}
\end{document}